\documentclass[fleqn,usenatbib]{mnras}

\usepackage{newtxtext,newtxmath}
\usepackage{graphicx}
\usepackage{caption}
\usepackage{subcaption}
\usepackage{placeins}
\usepackage{multirow}

\usepackage[T1]{fontenc}
\usepackage{ae,aecompl}
\usepackage{lineno}
\usepackage{listings}

\usepackage{graphicx}
\graphicspath{ {./images/} }

\usepackage{amsmath}	

\usepackage{layout}

\title[Faster BAO Fitting]{Accelerating BAO Scale Fitting Using Taylor Series}
\author[Hansen et al.]{
Matthew T. Hansen,$^{1}$ \thanks{E-mail: matthew.hansen@ufl.edu (MH)}
Alex Krolewski,$^{2, 3}$ \&
Zachary Slepian$^{1, 4}$
\\
$^{1}$Department of Astronomy, University of Florida, Gainesville, Florida 32611, USA \\
$^{2}$AMTD Fellow, Waterloo Centre for Astrophysics, University of Waterloo, Waterloo ON N2L 3G1, Canada \\
$^{3}$Perimeter Institute for Theoretical Physics, 31 Caroline St. North, Waterloo, ON N2L 2Y5, Canada
\\
$^{4}$Lawrence Berkeley National Laboratory, Berkeley, CA 94709, USA
}

\date{Accepted XXX. Received YYY; in original form ZZZ}

\pubyear{2021}

\begin{document}
\label{firstpage}
\pagerange{\pageref{firstpage}--\pageref{lastpage}}
\maketitle






\begin{abstract}
\noindent The Universe is currently undergoing accelerated expansion driven by dark energy. Dark energy's essential nature remains mysterious: one means of revealing it is by measuring the Universe's size at different redshifts. This may be done using the Baryon Acoustic Oscillation (BAO) feature, a standard ruler in the galaxy 2-Point Correlation Function (2PCF). 
In order to measure the distance scale, one dilates and contracts a template for the 2PCF in a fiducial cosmology, using a scaling factor $\alpha$.
The standard method for finding the best-fit $\alpha$ is to compute the likelihood over a grid of roughly 100 values of it. This approach is slow; in this work, we propose a significantly faster way. Our method writes the 2PCF as a polynomial in $\alpha$ by Taylor-expanding it about $\alpha = 1$, exploiting that we know the fiducial cosmology sufficiently well that $\alpha$ is within a few percent of unity. The likelihood resulting from this expansion may then be analytically solved for the best-fit $\alpha$. 
Our method is 48--85$\times$ faster than a directly comparable
approach in which we numerically minimize $\alpha$, and $\sim$12,000$\times$ faster than the standard iterative method.
Our work will be highly enabling for  upcoming large-scale structure redshift surveys such as that by Dark Energy Spectroscopic Instrument (DESI).
\end{abstract}



\section{Introduction}
\label{sec:intro}

In the early, radiation-dominated universe, primordial overdensities in the relativistic plasma grew gravitationally and were opposed by the radiation pressure of the photons, creating spherical sound waves \citep{Sakharov66, Peebles_Yu, Sunyaev_1970, Bond_84}. These acoustic oscillations are a powerful cosmological probe, both in
the Cosmic Microwave Background (CMB) anisotropies and in the Baryon Acoustic Oscillation (BAO) feature in the large-scale distribution of galaxies \citep{Eisentein_98,  Linder03, Slepian16}. 
At recombination, the Universe becomes neutral and therefore transparent to photons. Slightly later, the photons decouple
from the baryons and the 
sound waves halt. Velocity overshoot (\textit{i.e.}\ that the density mode that grows to late times follows the spatial structure of the  velocity at decoupling; \citealt{Vishniac80}) leaves a sharp bump in the baryon density Green's function \citep{Slepian16}. 
Over time, matter collects around both the initial overdensity that sourced the sound wave,
and the distant bump. The baryonic feature is a small perturbation compared to the central peak because the dark matter fraction is considerably larger than the baryon fraction \citep{Bashinsky01,Bashinsky02}.
This BAO feature, whose scale is given by the sound horizon at the drag epoch, $\sim$150 Mpc,
offers a standard ruler. 

Over the past 30 years, observations of BAO in both the CMB and the distribution of galaxies have led to percent-level measurements of the composition of the Universe. Moreover, they have cemented $\Lambda$CDM (a cosmological constant for dark energy, plus Cold Dark Matter) as the standard model of cosmology \citep{WMAP7,WMAP9,Planck18,Planck18Params,Alam21}.

The BAO feature in galaxy clustering is primarily useful as a standard ruler
\citep{EisensteinHuTegmark,BlakeGlazebrook,SeoEisenstein}.
Assuming no extra components in the early Universe, the sound horizon
is determined extremely precisely by CMB observations, with \cite{Planck18} providing a precision of 0.2\%.  Therefore, observations of the large-scale galaxy correlation functions at different redshifts allow an inference of cosmological distances (\textit{e.g.} \citealt{Weinberg13}).  This is complementary to the standard candles provided by type Ia supernovae, which enabled the first detection of the accelerating expansion of the Universe caused by dark energy \citep{Riess98,Perlmutter99}. Thus, precise measurements of the expansion history from BAO can continue to probe the time-variability of dark energy,
with $w$, the dark energy equation of state, constrained to being within 3\%
of a cosmological constant by \citet{Alam21}.

The BAO feature was first clearly detected in the final release
of the Two-degree-Field Galaxy Redshift Survey (2dFGRS) \citep{Cole05} and the Sloan Digital Sky Survey (SDSS) Data Release 3 (DR3) \citep{Eisenstein05}, though hints had been seen earlier \citep{Percival01}.  The BAO analyses of the DR7 SDSS Luminous Red Galaxy (LRG) and main galaxy samples established the BAO as a robust probe of cosmology \citep{Percival07, ESW07}. Subsequent analyses extended it to higher redshifts
 in the WiggleZ survey \citep{Blake11};
 $z \sim 0.1$ in the Six-degree-Field Galaxy Survey (6dFGS) \citep{Beutler11};
 improved precision by using reconstruction to partially
 reverse nonlinear smearing of the BAO peak \citep{Eisenstein07, Padmanabhan12};
 and much higher redshifts ($z$ $\sim$ $2.5$) in the Ly-$\alpha$
 forest \citep{Busca13, Slosar13, FontRibera14}. The BAO was first detected in the 3-Point Correlation Function (3PCF) by \cite{Slepian_2017} and in the bispectrum by \cite{PearsonSamushia}; it has also now been detected in cosmic voids \citep{Kitaura16}.
 The SDSS Baryon Oscillation Spectroscopic Survey (BOSS) was designed specifically to target
  a large sample of $z$$\sim$$0.5$ LRGs
  for BAO measurements and yielded a $1-2\%$ distance
   measurement at $z = 0.32$ and $z = 0.57$ \citep{Alam17}.
   Subsequently, the extended BOSS survey (eBOSS) extended the LRG sample of BOSS to $z$$\sim$$0.72$,
   and added BAO measurements from Emission Line Galaxies (ELGs) and quasars
   at $1 < z < 2$ to fill in the gap between the galaxies and the Ly-$\alpha$
   forest \citep{Alam21}.
Ongoing and future surveys such as that by Dark Energy Spectroscopic Instrument (DESI), Euclid, and the Nancy Grace Roman Space Telescope (NGRST) promise considerably better precision. For instance, DESI
should achieve a 0.5-1\% precision measurement
of the distance scale in bins of $\Delta z = 0.1$ \citep{Aghamousa16}, and Euclid \citep{Laureijs11,Euclid20} and NGRST \citep{Spergel15} augur similar constraining power.

The BAO feature contains the most robust
information about the cosmological parameters
from galaxy clustering statistics.
Some early studies directly
fit $\Lambda$CDM models with varying parameters
to the galaxy correlation function or power spectrum \citep{Tegmark06}. On the other hand, other works
instead fit a dilation parameter to the BAO feature and marginalize over the broadband shape
\citep{Eisenstein05}.
This isolates the distance information, which is the most robust part of the BAO measurement, from the broadband shape,
which is more challenging to model.

\begin{figure}
    \includegraphics[width=8.5cm]{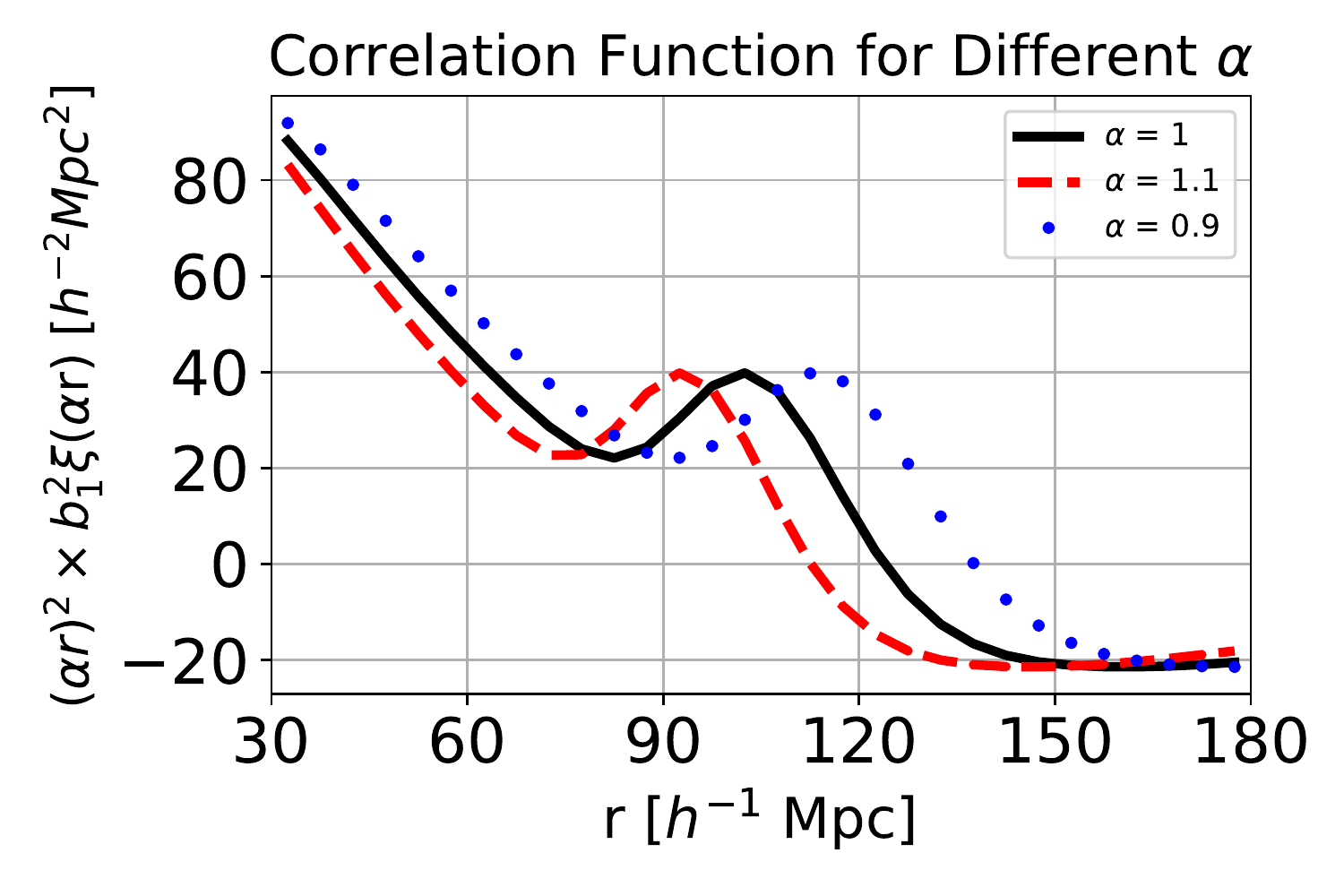}
    \caption{The scaling parameter $\alpha$ encodes the impact of moving the BAO feature to larger or smaller physical distances. In this plot, we use the fiducial value of linear bias $b_1 = 2.42$. $\alpha = 1$ (solid black curve) means no scaling is needed, $\alpha < 1$ (dotted blue curve) means that the true BAO feature is at larger scales than in the template, and $\alpha > 1$ (dashed red curve) means that the true feature is at smaller scales. Recalling that the 2PCF is defined as the overdensity of pairs of galaxies in the survey at a given separation, a larger value of $\alpha$ corresponds to galaxies' being closer together than expected. This implies that they are further away than in the fiducial cosmology (in the same way that an object of fixed transverse length appears smaller if further away).}
    \label{fig:alpha_scaling}
\end{figure}

Subsequently, this dilation method developed
into a standard method to determine
the BAO scale \citep{Xu2012,Anderson12,Anderson14,Ross15}.
In this method, a correlation function
or power spectrum template is produced at a
fixed cosmology, and the data is used to constrain the dilation parameter, $\alpha$, that
stretches or squeezes the template; we show this schematically in Figure \ref{fig:alpha_scaling}.
The dilation is different in the directions perpendicular to and parallel to the line of sight, producing
a characteristic distortion of the BAO feature
called the Alcock-Paczy\'nski effect if one uses the wrong cosmology \citep{AlcockPaczynski}.
Thus often two dilation parameters are used, $\alpha_{\parallel}$ and $\alpha_{\perp}$,
and they are sometimes referred to as the Alcock-Paczy\'nski parameters.\footnote{Although sometimes only a single isotropic parameter $\alpha$ is fit due to the low signal-to-noise of the  measurement \citep{Ross15,Carter18,deMattia20}.}
The result of the BAO fitting
is a best-fit value of $\alpha$ and and the error on $\alpha$ (or $\alpha_{\parallel}$ and $\alpha_{\perp}$ and their covariance matrix), and subsequently
cosmological parameters are fit using $\alpha$ and its covariance matrix.
Often this method is
referred to as the ``compressed statistics'' approach (the other piece of it is $f\sigma_8$, from Redshift Space Distortions (RSD), with $f$ the logarithmic derivative of the linear growth factor and $\sigma_8$ the rms density field fluctuations on 8 $h^{-1}$ Mpc scales).
This approach implicitly assumes a prior
(from existing measurements) on the shape of the power spectrum, typically determined from the best-fit parameters from the CMB.
This is a reasonable approach because the CMB constraints on the power spectrum
shape are far more powerful than the
large-scale structure constraints.

Recently, advances in perturbation theory (PT)
modelling for the broadband power spectrum
have led to renewed interest in directly
fitting $\Lambda$CDM parameters to the power
spectrum or correlation function, bypassing the
compressed statistics approach \citep{Ivanov20,Damico20,Philcox20,Chen21}.
However, the compressed statistics
approach remains valuable for a variety of reasons:
it is model-independent and provides
 a straightforward dataset to which to compare alternative cosmological models;
 it separates information from the early
 Universe and late Universe; and the
 compressed likelihood has a much smaller
 data vector and is easier to handle.
 As a result, the compressed
 statistics will continue to play an important
 role in future analyses,
 and we focus on accelerating their
inference from the clustering statistics in this work.




The standard approach for finding $\alpha$ is to to produce templates over a wide range (typically $0.8$ to $1.2$) of $\alpha$, compute the log likelihood of each template, and select $\alpha$ with the highest likelihood or the lowest $\chi^2$ (\textit{e.g.}\ as in the \textsc{BAOFit} software of \citealt{Ross15}).\footnote{ \url{https://github.com/ashleyjross/BAOfit}} On average, it takes about 2 to 3 seconds to find the optimal $\alpha$ on a typical CPU. 
This step may need to be repeated both across many thousands of mocks and for different analysis choices. Other steps in the BAO analysis chain (pair-counting for the 2-Point Correlation Function, or fitting cosmological parameters to the distances in an MCMC chain) may be much more costly, but would not necessarily have to be repeated many times. We believe that it is best to optimize the speed of the analysis at every part of the process possible.

In this work, we present a more efficient analytic method for obtaining the optimal $\alpha$.
Our method is $45$--$85$ $\times$ faster than numerically
minimizing $\chi^2$ as a function of $\alpha$ using the 
same 2PCF model that our method uses (but without the Taylor expansion), or $\sim$12,000 times faster than the standard method
from \textsc{BAOFit} iterating over many values of $\alpha$, which is not perfectly comparable as it does use a slightly more complicated 2PCF model.

Our method uses Taylor series to achieve this speedup. We start by calculating the 2PCF including only linear bias, and then Taylor-expand that model about $\alpha$ = 1 (taking advantage of the fact that the best-fit cosmology is known quite precisely and hence $\alpha$ will not be very far from this value). In \S\ref{sec:model_defn} we construct mock data and define our Taylor series model. We choose to simplify our model and not include the additional polynomial terms of \textit{e.g.}\ \cite{Ross15}. In \S\ref{sec:fix_bias} and \S\ref{sec:marg_bias} we show two cases: one with a fixed linear bias, and one where we marginalize over the linear bias. In \S\ref{sec:performance}, we compare the timing of our analytic method to both that of the standard method as well as that of a numerical method to minimize $\chi^2$ in the same model that we use.\footnote{The repository for implementation of our method: \url{https://github.com/matthew3hansen/BAO_Scaling_Taylor_Series}} 
We also discuss the range of validity for the analytic method, quantifying how much deviation is allowed before it produces a significant systematic error on upcoming constraints.
We conclude in \S\ref{sec:conclusions}. In Appendix \ref{sec:poly} we outline how to marginalize over additional polynomial terms in the model, designed to remove any broadband effects; we leave implementation for future work.

\section{Defining The 2PCF Model and Mock Data}
\label{sec:model_defn}

We create a mock dataset
roughly mimicking the properties
of the DESI survey \citep{Levi13,Aghamousa16}
to test our approach.
We start by generating a linear
power spectrum using \textsc{CCL}\footnote{\url{https://github.com/LSSTDESC/CCL}} \citep{Chisari19}, which in turn calls \textsc{CLASS} \citep{Class} with densities (in units of the critical density) for the matter of
$\Omega_{\rm m} = 0.315$, for the  baryons  $\Omega_{\rm b}$ = 0.045,  $h \equiv H_0/(100\;{\rm km/s/Mpc}) = 0.67$, primordial amplitude of fluctuations $A_s$ = 2.1 $\times$ $10^{-9}$, and scalar spectral tilt $n_s$ = 0.96.
To create a correlation function with realistic behavior
around the BAO scale, we model the galaxy power spectrum $P_{\rm gg}$ as a linear bias times $b$ the infrared-resummed
power spectrum $P_{\rm IR-resum}$ at wavenumber $k$ and redshift $z$:
\begin{equation}
    P_{\rm gg}(k,z) = b_1^2 P_{\textrm{IR-resum}}(k,z). \label{eqn:power}
\end{equation}
Infrared resummation allows for more accurate modelling
of the bulk displacements in the BAO feature than does standard perturbation theory \citep{Senatore15}.
The infrared-resummed power spectrum
is generated from the linear power spectrum using the \textsc{FASTPT} code \citep{McEwen16,Fang17}.

We work at $z = 0.7$, a representative
redshift for the DESI LRG sample.
This galaxy sample has $b_1 = 1.7/D(z=0.7) = 2.42$, where $D$ is the linear growth factor.
This form for the bias
evolution is typically
assumed for DESI LRGs \citep{Aghamousa16} and roughly matches early clustering data \citep{Kitanidis}.
For the purposes of constructing
data to test the method, we 
drop higher-order biases.
We then inverse Fourier-transform the galaxy power spectrum $P_{\rm gg}(k)$ to the correlation function, $\xi(r)$,
and work with the correlation function throughout this paper. We have
\begin{equation}
    \xi_{\rm gg}(r) = \int \frac{k^2 dk}{2\pi^2} j_0(kr) P_{\rm gg}(k),
\end{equation}
where $j_0(x) = \sin x/x$ is the spherical Bessel function of order zero. 

\begin{figure*}
\centering
     \begin{subfigure}[b]{0.49\textwidth}
        \centering
         \includegraphics[width=9.5cm]{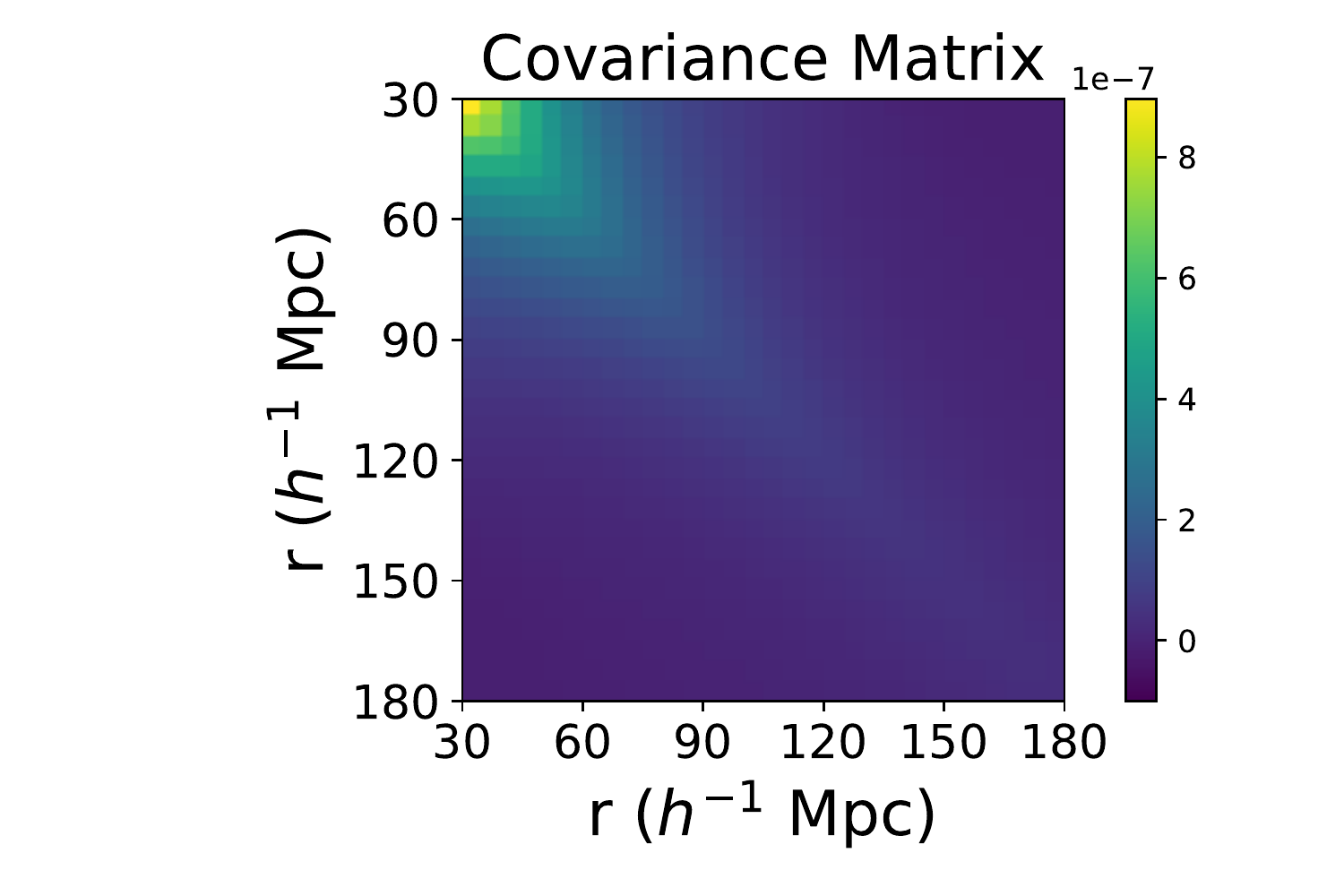}
     \end{subfigure}
     \begin{subfigure}[b]{0.49\textwidth}
        \centering
         \includegraphics[width=9.5cm]{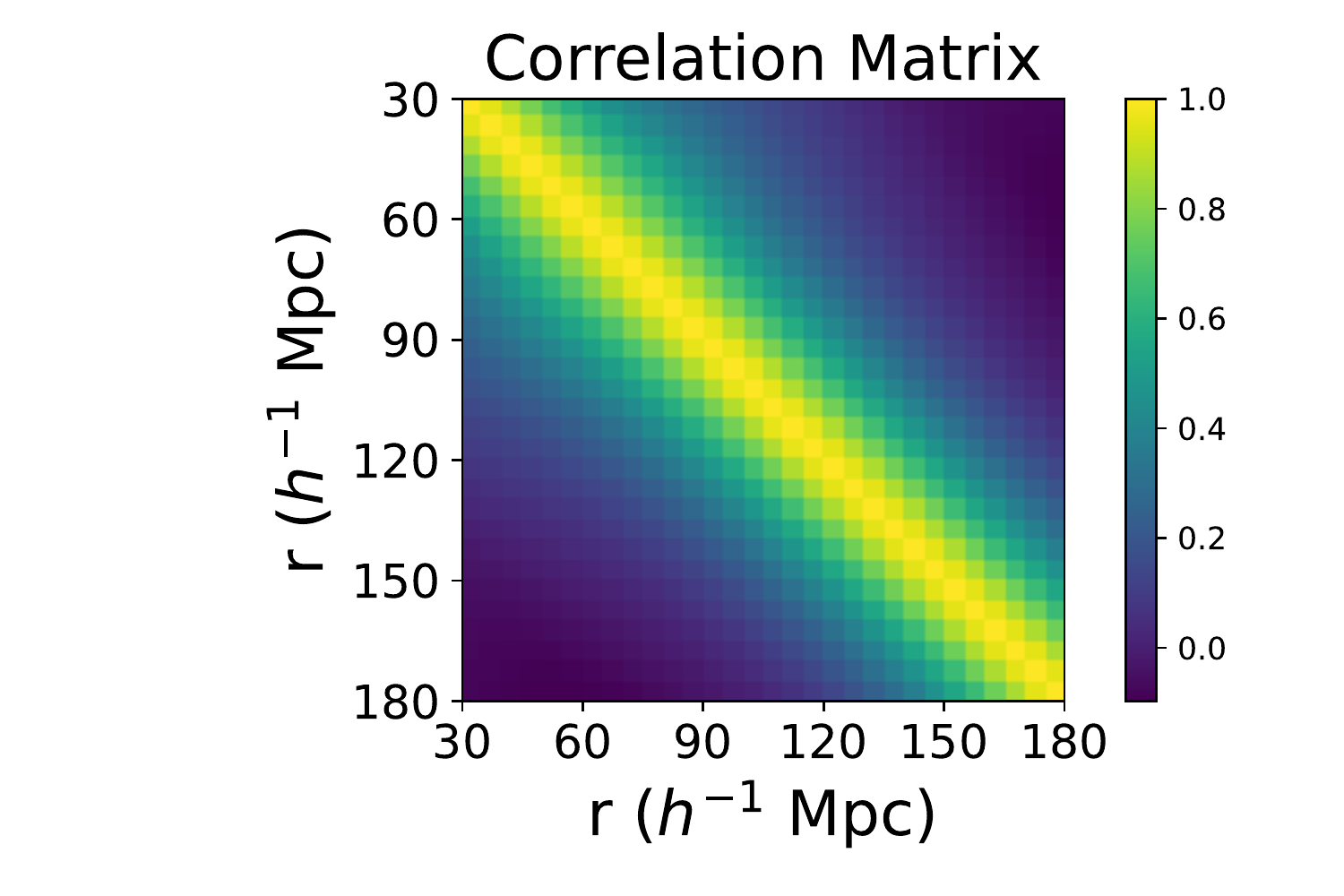}
     \end{subfigure}
     \begin{subfigure}[b]{0.6\textwidth}
         \includegraphics[width=10cm]{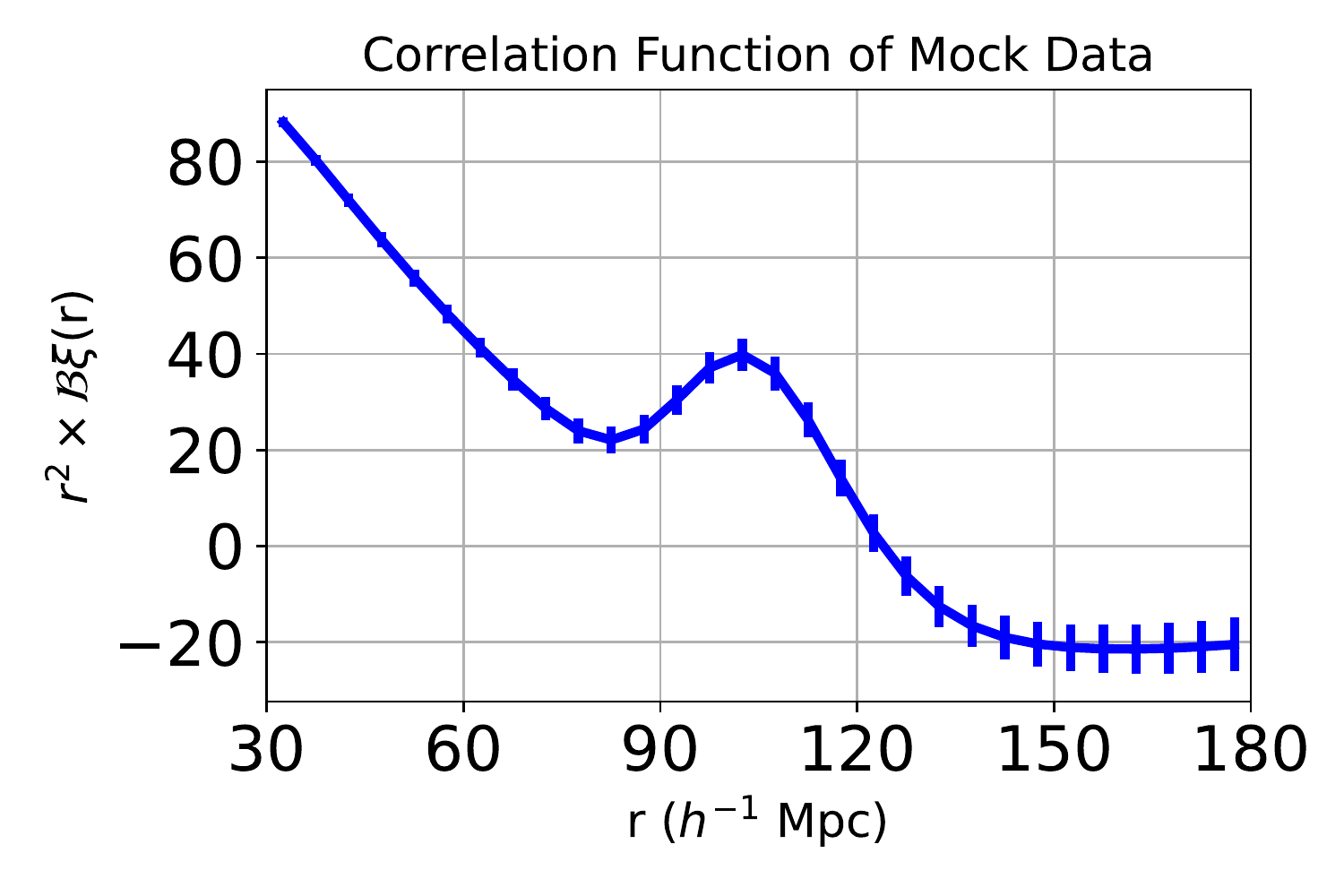}
     \end{subfigure}
     \caption{\textit{Top left:} The analytic covariance matrix for the 2PCF, computed from equation (\ref{eqn:analytic_cov}), at redshift 0.7 and a number density of $6 \times 10^{-4}$ $h^3$ Mpc$^{-3}$. \textit{Top right:} The correlation matrix, with elements $C_{ij}/\sqrt{C_{ii} C_{jj}}$, computed from the analytic covariance matrix. We note that the colorbar goes slightly below zero for both the covariance and correlation matrix (\textit{i.e.} some elements are slightly negative). \textit{Bottom:} The correlation function on a mock data set with $\alpha$ = 1, $b_1 = 2.42$ ($\mathcal{B} \equiv b_1^2$ = 5.88), and error bars from the analytic covariance matrix.}
     \label{fig:data_and_covariance}
\end{figure*}

We use the analytic expression for the correlation function's covariance matrix \citep{Cohn06},
\begin{align}
    \textrm{Cov}(\xi(r),\, \xi(r')) &= \frac{1}{V_{\rm eff} \pi^2} \int k^2 dk j_0(kr) j_0(kr') P(k)^2 \nonumber \\
    &+ \frac{2}{V_{\rm eff} \bar{n} \pi^2} \int k^2 dk j_0(kr) j_0(kr') P(k) \nonumber \\
    &+ \frac{1}{V_{\rm eff} \bar{n}} \frac{2}{4 \pi r^2} \delta_{\rm D}^{[1]}(r-r')(1 + \xi(r)) \label{eqn:analytic_cov}.
\end{align}
$\delta_{\rm D}^{[1]}$ is the 1D Dirac delta function. The first two terms above are produced by the Gaussian density field and the final term by the shot noise
in the galaxy distribution.
$V_{\rm eff}$ is the effective
volume, related to the survey volume $V_{\rm survey}$
by
\begin{equation}
    V_{\rm eff} = \left( 1 + \frac{1}{\bar{n}P_{\rm eff}} \right)^{-2} V_{\rm survey}.
\end{equation}
The survey volume
is taken as the volume of a spherical
shell between $z = 0.6$ and $z = 0.8$, and the number density is $\bar{n} = 6 \times 10^{-4}$ $h^{3}$ Mpc$^{-3}$, appropriate
for the DESI LRG sample
\citep{Zhou_2020}. The effective anisotropic
power spectrum $P_{\rm eff}$ ({\textit i.e.}\ power spectrum at roughly the average wavenumber $k$ and angle to the line of sight $\mu$) is computed at $k = 0.14$ $h$ Mpc$^{-1}$ and $\mu = 0.6$ \citep{Aghamousa16}
and is given by
\begin{equation}
    P_{\rm eff} = \left(b_1 + f (\mu)^2\right)^2 P_{\rm lin}(k, z).
\end{equation}
Our fiducial
binning consists of 30 bins
with width $\Delta r = 5$ $h^{-1}$ Mpc between $r = 30$ $h^{-1}$ Mpc
and $r = 180$ $h^{-1}$ Mpc.
We average the covariance
over the bins, and show the fiducial
data and binned covariance in Figure~\ref{fig:data_and_covariance}. In Figure \ref{fig:L}, we show the log likelihood of the Taylor expanded model compared to the standard model. In Figure \ref{fig:fixed_b1} and Figure \ref{fig:marginal_b1}, we produce 100 different mock data sets, each with a different value of $\alpha$ corresponding to the labeled ``True $\alpha$.'' 

We also specifically choose not to add noise into the data that we use to test the best-fit $\alpha$ in the Taylor series model. 
Adding noise would introduce error into our Taylor series model, since the best-fit $\alpha$ would be equal to the true $\alpha$ plus a small random number.  This noise-produced offset in $\alpha$ would interfere with our aim of comparing to the input $\alpha$ specifically to ensure that our method's Taylor expansion retain the desired accuracy in the recovered $\alpha$.
However, in \S \ref{sec:noisy_comp} and \S \ref{sec:performance} we perform several tests to check the robustness of our method in the presence of noise.

Since we know the value of $\alpha$ is close to unity, we can Taylor-expand $\xi(\alpha r)$ about $\alpha = 1$ as:
\begin{align}
    \xi(\alpha r) &= \xi (r) + \xi' (r)\Delta \alpha + \frac{1}{2} \xi''(r)\Delta \alpha^2 + \frac{1}{6} \xi'''(r) \Delta \alpha^3,
    \label{eqn:poly_xi}
\end{align}
\noindent where $\Delta\alpha \equiv \alpha - 1$ and prime denotes the partial derivative with respect to $\alpha$. We Taylor-expand to third order in $\alpha$, which allows for a consistent
expansion to second order in the log likelihood, as will be seen.
We show in \S\ref{sec:fix_bias} that a Taylor expansion to second order is insufficient to achieve the necessary accuracy in $\alpha$.

A similar Taylor expansion was presented in \cite{xu_2013}, though they expanded both the isotropic
and anisotropic parts of the BAO scaling, and kept only first-order terms. Our work expands on this method,
and shows that to achieve sufficient accuracy
for current-generation surveys, we must expand to third order rather than first order.


\subsection{Template Derivatives}


Since we Taylor-expand our model to third order, we require the first, second and third derivatives of $\xi(\alpha r)$ with respect to $\alpha$, evaluated at $\alpha = 1$.

To enable efficiently taking these derivatives without resorting to numerical differencing, which can suffer from stability issues, we begin with our definition of $\xi(\alpha r)$. We have:
\begin{align}
    \xi(\alpha r) = \int \frac{k^2 dk}{2\pi^2} j_0(k \alpha r) P_{\rm lin}(k).
\end{align}
Taking $\partial/\partial \alpha$, we find:
\begin{align}
    \frac{\partial}{\partial \alpha} \xi(\alpha r) = \int \frac{k^2 dk}{2\pi^2} \frac{\partial}{\partial \alpha} \big[j_0(k \alpha r) \big] P_{\rm lin}(k).
\end{align}
We use the recursion relation for the spherical Bessel functions: \footnote{\url{https://dlmf.nist.gov/10.51}}
\begin{align}
\label{eqn:rec}
    j_n'(z) &= -j_{n + 1}(z) + \frac{n}{z}j_n(z), \qquad n = 0,1,2,3\ldots
\end{align}
We find that

\begin{align}
    \frac{\partial}{\partial \alpha}\xi(\alpha r) \bigg|_{\alpha = 1} = -r \int \frac{k^2 dk}{2\pi^2} k j_1(kr) P_{\rm lin}(k).
\end{align}
Applying equation (\ref{eqn:rec}) again to obtain the second derivative, we find
 \begin{align}
     &\frac{\partial^2}{\partial \alpha^2} \xi(\alpha r)\bigg|_{\alpha = 1}\\
     &\qquad \qquad \qquad= r^2 \int \frac{k^2 dk}{2\pi^2} k^2 \left[ j_2(kr) - \frac{1}{kr} j_1(kr)\right] P_{\rm lin}(k).\nonumber
 \end{align}
The third derivative is then:
\begin{align}
    \frac{\partial^3}{\partial \alpha^3} \xi(\alpha r)\bigg|_{\alpha = 1} &= r^3 \int \frac{k^2 dk}{2\pi^2} k^3 \left[ - j_3(k r) + \frac{2}{k r} j_2(k r)\right] P_{\rm lin}(k) \nonumber \\
    &+ r \int \frac{k^2 dk}{2\pi^2} k j_1(k r) P_{\rm lin}(k) \\
    &+ r^2 \int \frac{k^2 dk}{2\pi^2} k^2 \left[ j_2(k r) + \frac{1}{k r} j_1(k r)\right] P_{\rm lin}(k).\nonumber
 \end{align}
With these templates in hand, we may now proceed to evaluate the Taylor expansion (\ref{eqn:poly_xi}) and then solve for the maximum-likelihood $\Delta \alpha$.
\section{Optimal $\alpha$ at Fixed Linear Bias}
\label{sec:fix_bias}
\begin{figure*}
     \centering
     \begin{subfigure}[b]{0.49\textwidth}
         \centering
         \includegraphics[width=\textwidth]{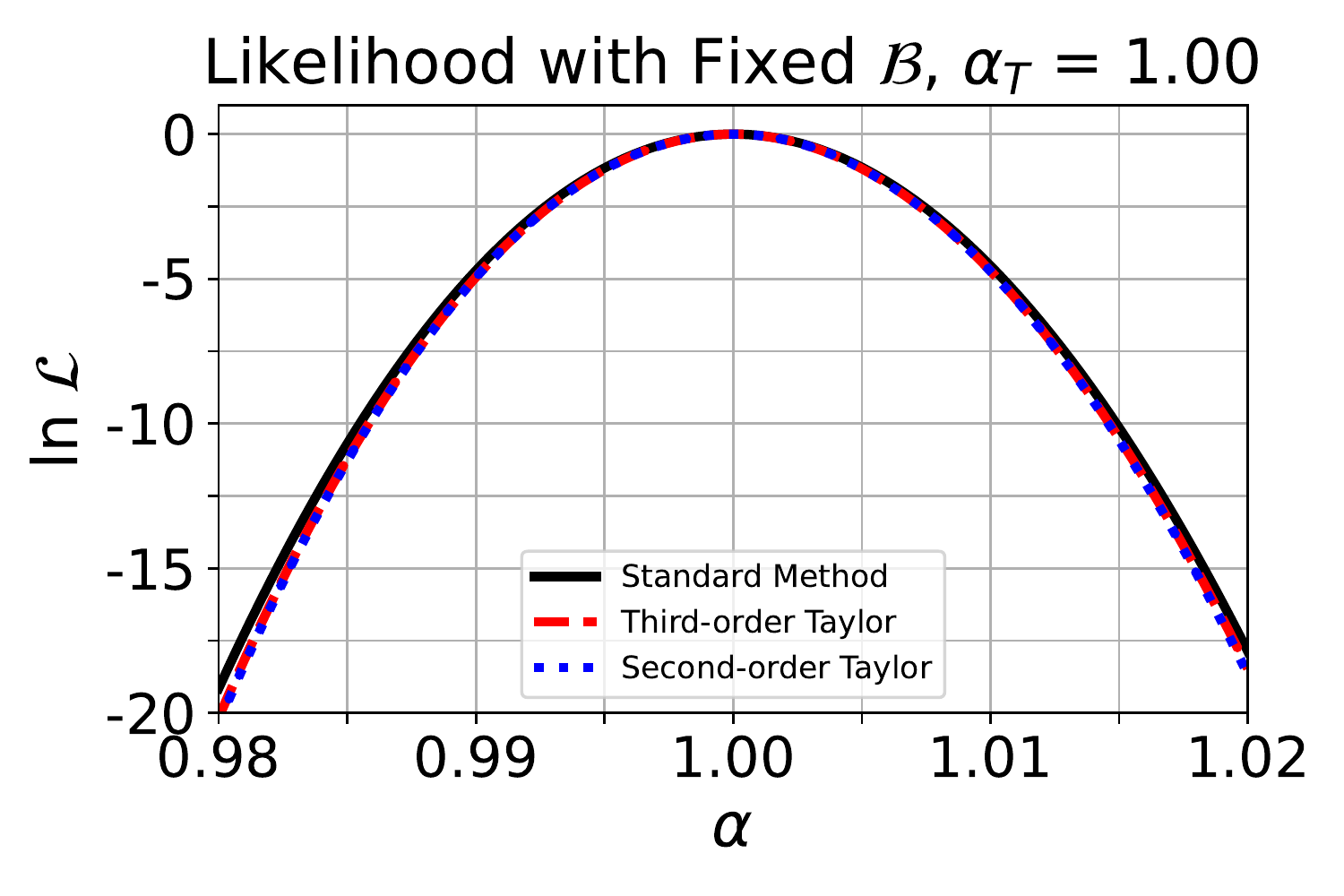}
     \end{subfigure}
     \begin{subfigure}[b]{0.49\textwidth}
         \centering
         \includegraphics[width=\textwidth]{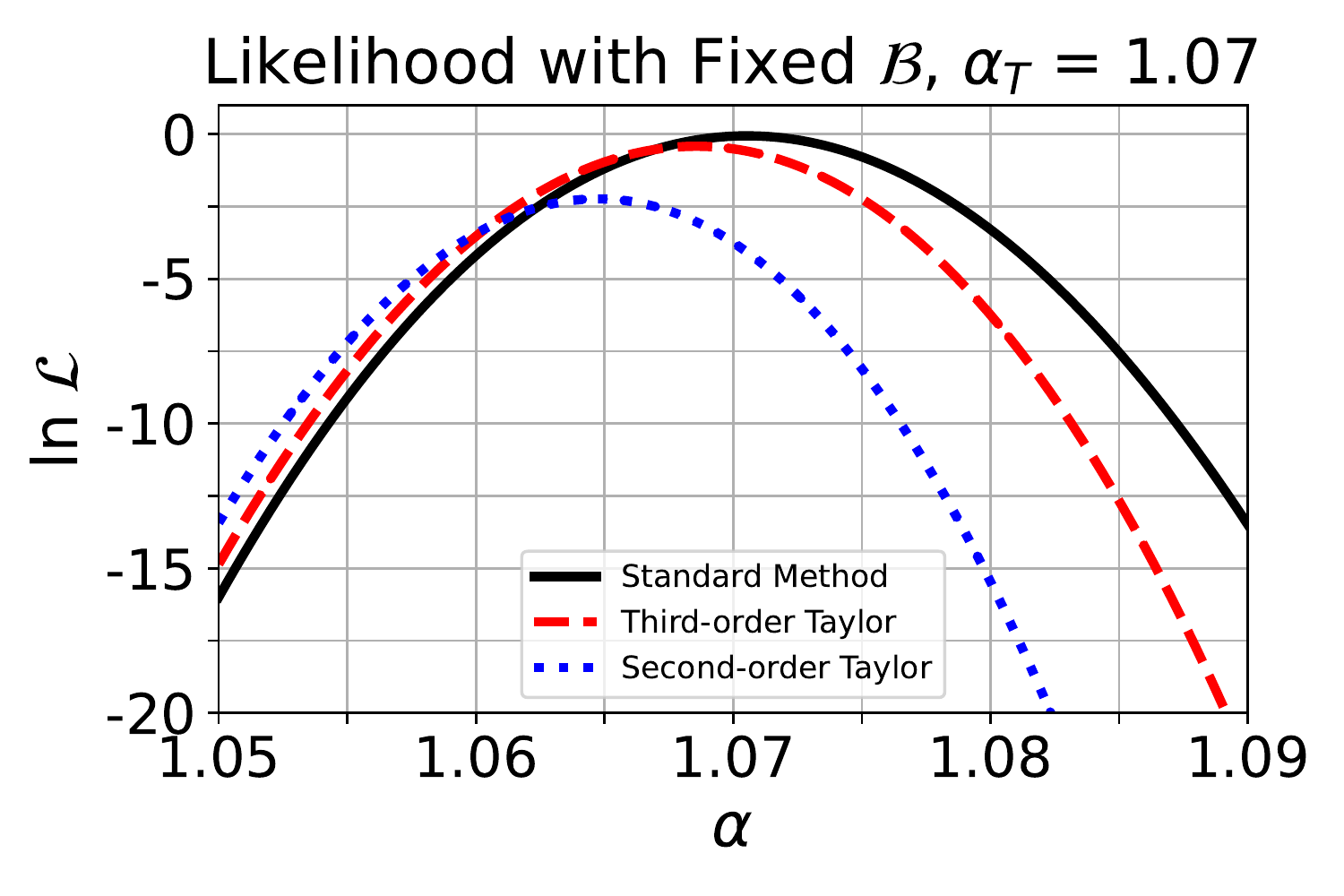}
     \end{subfigure}
     \begin{subfigure}[b]{0.5\textwidth}
         \centering
         \includegraphics[width=\textwidth]{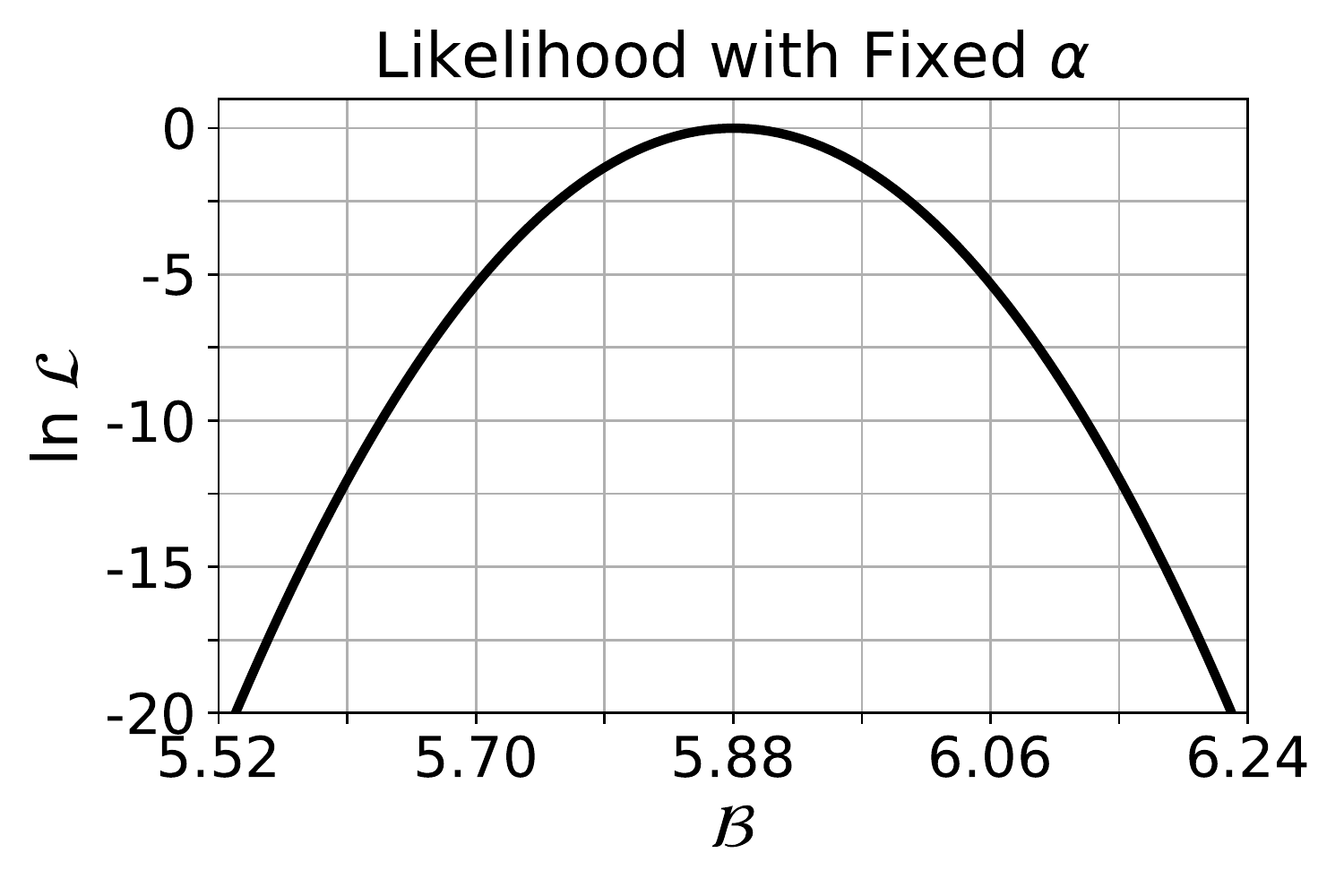}
     \end{subfigure}
     \begin{subfigure}[b]{0.49\textwidth}
         \centering
         \includegraphics[width=\textwidth]{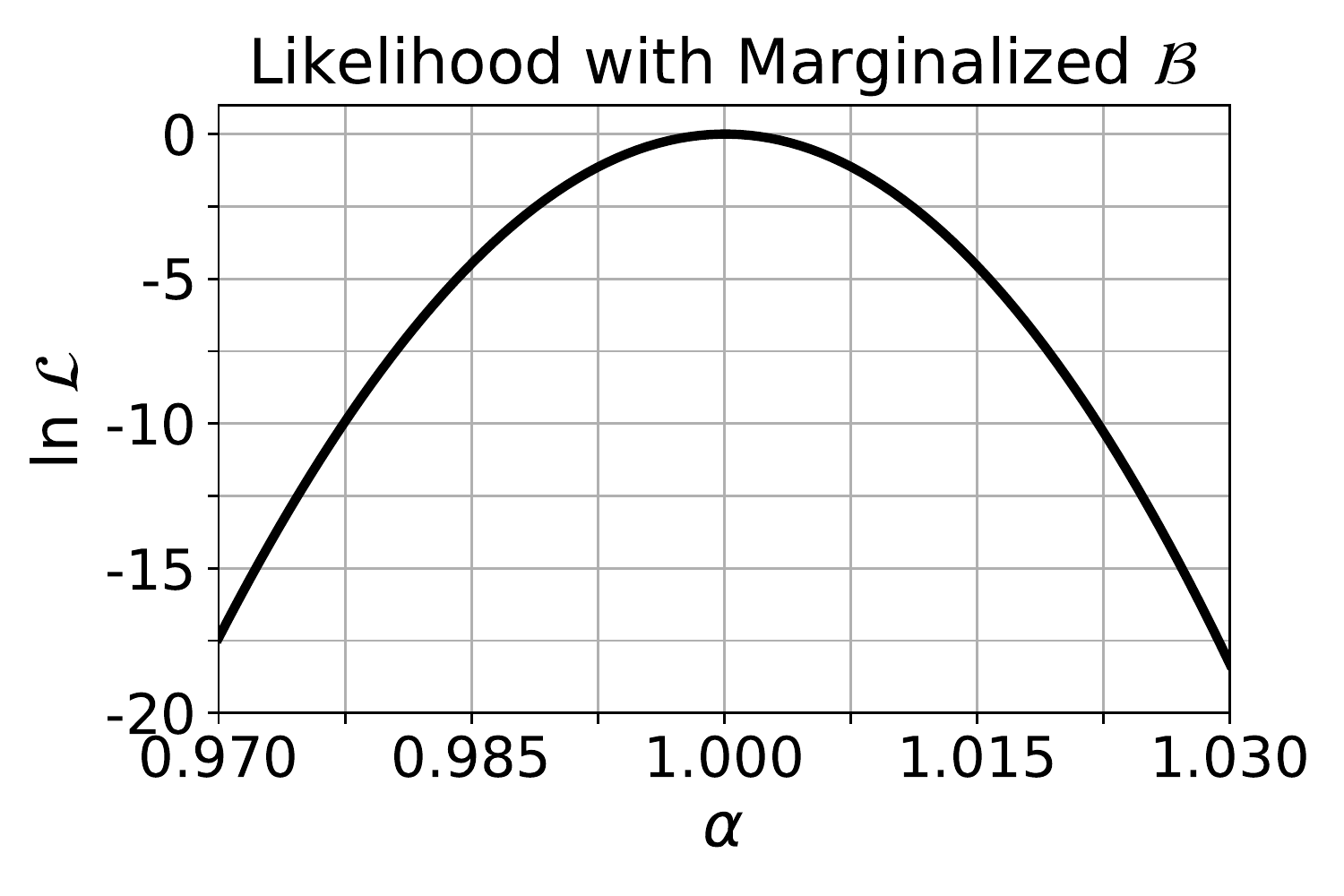}
     \end{subfigure}
        \caption{\textit{Top left:} The likelihood for the second-order and third-order Taylor expansions (respectively dotted blue and dashed red)  agree extremely well with the full model (solid black) with the mock data scaled by $\alpha = 1.0$. The standard method (solid black) is one where $\alpha$ is varied with a spline interpolation. \textit{Top right:} At $\alpha = 1.07$, the second-order Taylor expansion has a large discrepancy in the log likelihood compared to both the standard method and to the third-order Taylor expansion. \textit{Bottom left:} The likelihood as a function of the bias squared, \textit{i.e.} $\mathcal{B} \equiv b_1^2$. The likelihood is peaked at the true value in the mock data ($b_1 = 2.42,\;\mathcal{B} = 5.88$). \textit{Bottom right:} The likelihood of the third-order Taylor expansion after marginalizing over $\mathcal{B}$, for a range of $\alpha$. }
        \label{fig:L}
\end{figure*}
We now turn our focus to the log likelihood of our data with respect to the model. 
To analytically obtain the optimal value for $\alpha$, we maximize the log likelihood by setting its derivative to zero. By retaining terms up to third order in $\Delta \alpha$, we will obtain a quadratic in $\alpha$ when taking the likelihood's derivative, which we can then solve analytically for the optimal $\Delta \alpha$.
Primes will denote the derivatives with respect to $\alpha$, $\vec{d}$ denotes the data vector, and $\vec{m}$ denotes the model vector. For these vectors, each element describes, respectively, the data and model correlation functions in a given radial bin. 

The log likelihood is:
\begin{align}
    -\ln \mathcal{L} &= 0.5 \left[\vec{d} - \vec{m}\right]{\bf C^{-1}} \left[\vec{d} - \vec{m}\right]^T,
\end{align}
where ${\bf C}$ is the covariance matrix.
We define $ \bf{C^{-1}}$ to have elements $\mathcal{P}^{lm}$, where $\mathcal{P}$ is the precision matrix. The superscripts  $l$ and $m$ identify the radial bin of each of the data or model vectors involved.
Using Einstein summation convention, the log likelihood becomes:
\begin{align}
    -\ln \mathcal{L} &= 0.5 [d_l - m_l] \mathcal{P}^{l m} [d_m - m_m].\label{eqn:likeli_compare}
\end{align}
Since all of these elements are now scalars, their order no longer matters. Hence we may rewrite the log likelihood as
\begin{align}
    -\ln \mathcal{L} &= 0.5 \mathcal{P}^{l m} [d_l d_m - d_l m_m - m_l d_m + m_l m_m]. \label{eqn:lnL}
\end{align}
Now, taking the derivative of the log likelihood with respect to $\Delta \alpha$, we find
\begin{align}
\frac{\partial}{\partial (\Delta \alpha)} \left[-\ln \mathcal{L}\right] &= 0.5\mathcal{P}^{lm}\big[ - d_l m_m' - m_l' d_m + m_l' m_m + m_l m'_m\big] \label{eqn:partial_like} \nonumber \\
&= 0.
\end{align}
Our model vector has elements $m_i$  (the value at the $i^{th}$ radial bin) as:
\begin{align}
    m_i = \mathcal{B} \xi(\alpha r_i),
    \label{eqn:model_label}
\end{align}
where $\xi (\alpha r)$ is defined in equation (\ref{eqn:poly_xi}) and $\mathcal{B} \equiv b_1^2$, with $b_1$ the linear bias. Figure \ref{fig:L} shows the log likelihood as a function of $\alpha$ and as a function of $\mathcal{B}$. These plots serve as our basis for approximating the model as a third-order Taylor series about $\alpha$.
Figure~\ref{fig:L} also argues that we need to Taylor expand
the model to third order rather than second order. At the point of expansion $\alpha = 1$, the difference between third and second order is negligible near the maximum log likelihood. However,
at $\alpha$ far from 1, third order is significantly more accurate than second order, as show in the upper right panel of Fig.~\ref{fig:L} at $\alpha = 1.07$.
Thus, going to third order significantly increases the range of $\alpha$ over which our method is accurate.

We now need to write out any terms in equation (\ref{eqn:partial_like}) involving the product of two models at different radial bins up to second order in $\Delta \alpha$. We substitute into equation (\ref{eqn:partial_like}) using equation (\ref{eqn:model_label}) for the model, revealing what degree in $\Delta \alpha$ each term contains. This substitution will allow us to group the terms in the equation into a quadratic in terms of $\Delta \alpha$:
\begin{align}
\frac{\partial}{\partial (\Delta \alpha)} \left[-\ln \mathcal{L}\right] = a(\Delta\alpha)^2 + b(\Delta\alpha) + c,
\label{eqn:18}
\end{align}
 with coefficients $a,\, b$ and $c$ as:
\begin{align}
    a &= \frac{1}{4} \mathcal{B} \mathcal{P}^{lm} \bigg[ \mathcal{B} \left( 3\xi'_{l} \xi''_{m} + 3\xi''_{l} \xi'_{m} + \xi'''_{l} \xi_{m} + \xi_{l} \xi'''_{m}\right) \nonumber \\
    &\qquad - d_l \xi'''_m - d_m \xi'''_l \bigg],
    \nonumber \\
    b &= \frac{1}{2}\mathcal{B} \mathcal{P}^{lm} \left[ \mathcal{B} \xi_{l} \xi''_{m} + \mathcal{B} \xi''_{l} \xi_{m} + 2 \mathcal{B} \xi'_{l} \xi'_{m} - d_l \xi''_{m} - d_m \xi''_{l}\right],
    \nonumber \\
    c &= \frac{1}{2}\mathcal{B} \mathcal{P}^{lm} \left[ \mathcal{B} \xi_{l} \xi'_{m} + \mathcal{B} \xi'_{l} \xi_{m} - d_l \xi'_{m} - d_m \xi'_{l}\right].
    \label{eqn:fixed_variable_labels}
\end{align}
We note here that $a,\, b$ and $c$ are scalars.

With these coefficients in hand, we may now solve the quadratic (\ref{eqn:18}) for $\Delta \alpha$ using the quadratic formula, as
\begin{align}
\Delta \alpha &= \frac{-b \pm \sqrt{b^2 - 4ac}}{2a}, \label{eqn:quadratic_delta_alpha}
\end{align}
with $a,b$ and $c$ defined as in equation (\ref{eqn:fixed_variable_labels}). The solution for $\Delta \alpha$, equation (\ref{eqn:quadratic_delta_alpha}), is only valid $\mathcal{B}$ is held constant. 

\begin{figure*}
    \begin{subfigure}[b]{0.49\textwidth}
        \includegraphics[width=8cm]{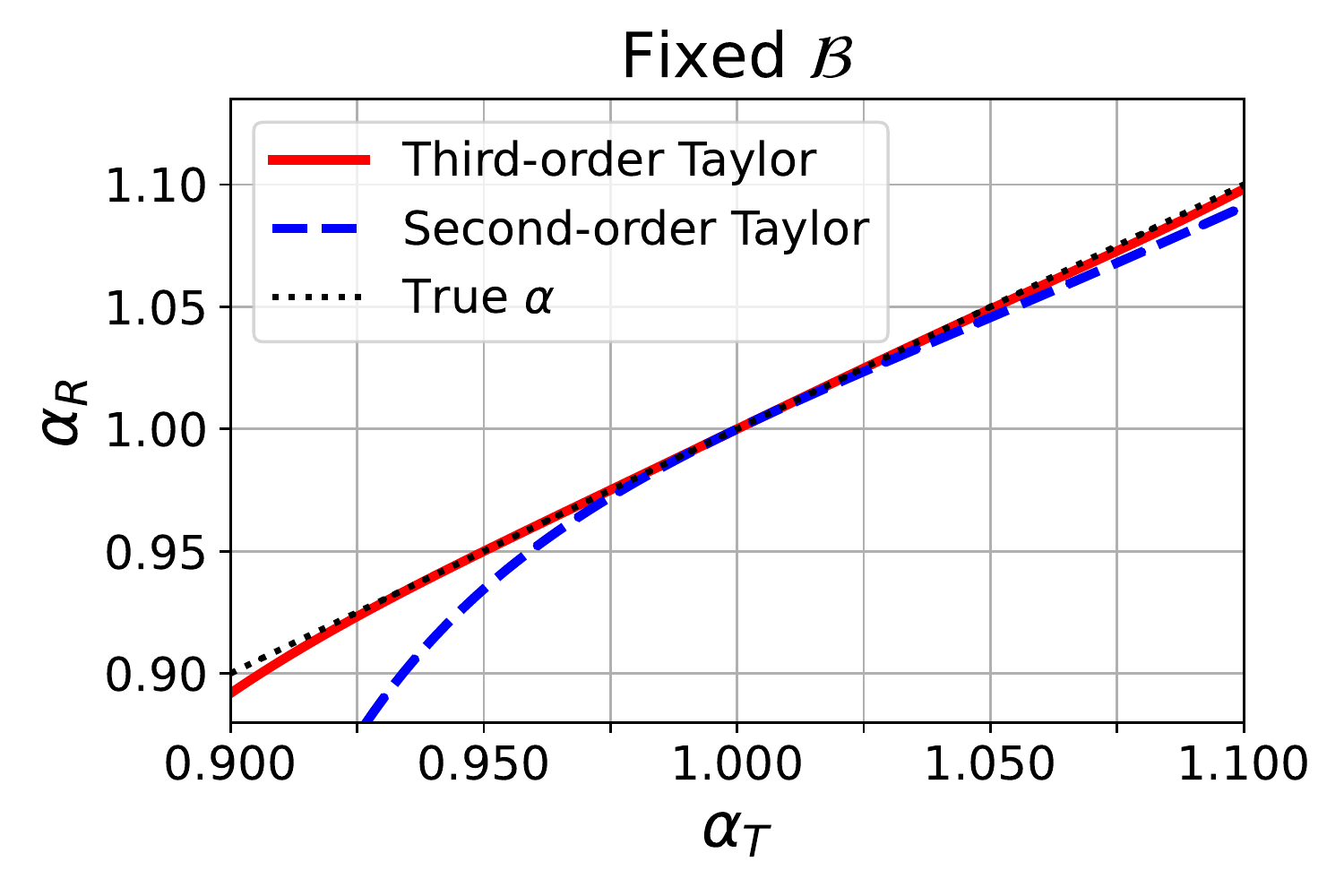}
    \end{subfigure}
    \begin{subfigure}[b]{0.5\textwidth}
        \includegraphics[width=8cm]{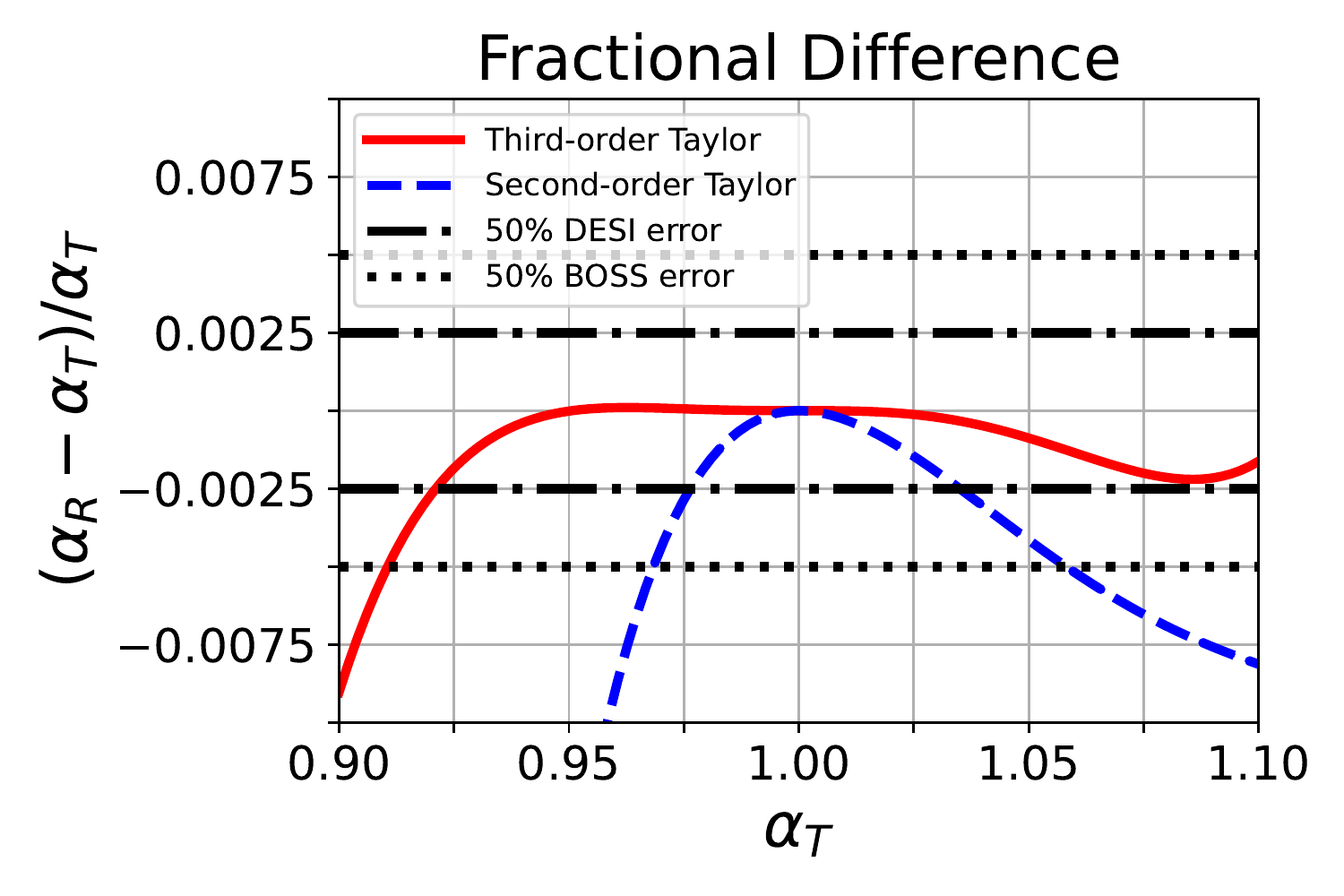}
    \end{subfigure}
    \caption{\textit{Left:} Comparison of the recovered $\alpha$, $\alpha_R$, of our fiducial method (third-order Taylor expansion; solid red) with the true inputted $\alpha$, $\alpha_T$,  (dotted black) as well as our method using the second-order Taylor expansion (dashed blue), which is considerably less accurate at large values of $\alpha$. The ``True $\alpha$'' line represents perfect recovery of the inputted $\alpha$.
    \textit{Right:} The fractional difference shows that our third-order method retains accuracy to $<0.5\%$ even if the true distance scale is 10\% different from the one used to generate the templates. We show in  dotted and dash-dotted lines where the error on $\alpha$ is 0.5\% and 0.25\%.  By comparison, BOSS achieves $\sim$1\% precision on $\alpha$, and DESI achieves $\sim$0.5\% precision.  This comparison shows that our method adds negligible systematic error relative to DESI's statistical precision over a range of almost 10\% in $\alpha$. Importantly, at present $\alpha$ is already known quite well (roughly one percent precision at $z$$\sim$$0.6$, a few percent at $z$$\sim$$1.5$), so we can center our Taylor expansion on a point that will be rather close---certainly well within $10\%$---to whatever true value DESI recovers.
    }
    \label{fig:fixed_b1}
\end{figure*}

In Figure \ref{fig:fixed_b1}, we hold $\mathcal{B}$ fixed to a value of 5.88 and create 100 different mock data sets, each with a different value of $\alpha$ ranging from 0.9 to 1.1 in steps of 0.002. We then compare the value equation (\ref{eqn:quadratic_delta_alpha}) returns for $\Delta \alpha$ to the inputted true value of $\Delta \alpha$. 
We note that the the recovered $\alpha$, $\alpha_{R}$, in Figure~\ref{fig:fixed_b1} is slightly different from the peak of the likelihood in the right panel of Figure~\ref{fig:L}. This is because
the likelihoods in Figure~\ref{fig:L} include the higher-order $\Delta \alpha^3$ and $\Delta \alpha^4$ terms, but these are omitted
in the Taylor series analytic solution.

This procedure's goal is to confirm that our method of finding $\Delta \alpha$ is accurate over a wide range of input $\Delta \alpha$. Our method is accurate to 0.7\% over an extremely wide range of $\alpha$, $0.9 < \alpha < 1.1$. This compares favorably to the errors on current and future surveys: BOSS constrains $\alpha$ to $\sim$1\% precision \citep{Alam17} and DESI will constrain $\alpha$ to $\sim$0.5\% precision \citep{Aghamousa16}. In Figure \ref{fig:fixed_b1}, we add horizontal lines corresponding to 50\% of the statistical error on these two surveys (\textit{i.e.}\ 0.25\% and 0.5\%) as a guide to determine the $\alpha$ at which our method stops working.  For 0.5\% errors on the distance scale, our method for the third-order Taylor expanded model works from $0.91 < \alpha < 1.16$. For 0.25\% distance errors, our method works from $0.92 < \alpha < 1.15$.
In contrast, the second-order Taylor-expanded model works to 0.5\% at $0.96 < \alpha < 1.055$, and 0.25\% at $0.975 < \alpha < 1.03$.
Moreover, the second-order Taylor-expanded model rapidly rises to very large errors at $\alpha_T < 0.95$. While the second-order
model may be accurate enough for DESI given that $\alpha_T$ will be quite close to one, we prefer the third-order model as it is far more robust away from $\alpha = 1$ (and therefore may work
better when, for instance, iterating over large numbers of mock data, some of which may have larger or smaller values of $\alpha$ just by chance).

This successful simplified case of finding the best fit $\Delta \alpha$ at a fixed $\mathcal{B}$ means we can now extend our approach to more complicated models. 
In the present section, we have treated $\mathcal{B}$ as a known constant, which is not entirely correct: typically the galaxy bias is unknown and marginalized over. Hence the present section serves as a proof of concept for our method. The next section, \S\ref{sec:marg_bias}, treats the more realistic case wherein one marginalizes over the unknown value of $\mathcal{B}$. 

\FloatBarrier

\section{Optimal $\Delta \alpha$ With Marginalization Over Linear Bias}
\label{sec:marg_bias}
We introduce marginalization over $\mathcal{B}$ as a means of accounting for $\mathcal{B}$ to mirror what is done in realistic BAO analysis, where the galaxy bias is unknown. Figure \ref{fig:L} shows the likelihood is peaked about the inputted $\alpha$ with marginalization over $\mathcal{B}$. We use the same definitions of $\mathcal{L}$, equation (\ref{eqn:lnL}), and $\vec{m}$, equation (\ref{eqn:model_label}). We write the likelihood as a quadratic in $\mathcal{B}$ to facilitate marginalization,
\begin{align}
    \mathcal{L}(\mathcal{B}) = \exp\left( -a \mathcal{B}^2 + b \mathcal{B} + c\right),
\label{eqn:likelihood_b}
\end{align}
with coefficients
\begin{align}
    a = \frac{1}{2} \mathcal{P}^{lm} &\bigg[ \xi_m \xi_l + \xi_m \xi'_l \Delta \alpha + \frac{1}{2} \xi_m \xi''_l \Delta \alpha^2 + \xi'_m \xi_l \Delta \alpha + \xi'_m \xi'_l \Delta \alpha^2 \nonumber \\
    &+ \frac{1}{2} \xi'_m \xi''_l \Delta \alpha^3 + \frac{1}{2}\xi''_m \xi_l \Delta \alpha^2 + \frac{1}{2} \xi''_m \xi'_l \Delta \alpha^3 + \frac{1}{4} \xi''_m \xi''_l \Delta \alpha^4 \nonumber \\
    &+ \frac{1}{6}\xi'''_m \xi_l \Delta \alpha^3 + \frac{1}{6}\xi'''_l \xi_m \Delta \alpha^3 \bigg], \nonumber \\
    \nonumber \\
    b = \frac{1}{2} \mathcal{P}^{lm} &\bigg[ d_l \left( \xi_m + \xi'_m \Delta \alpha + \frac{1}{2} \xi''_m \Delta \alpha^2  + \frac{1}{6}\xi'''_m \Delta \alpha^3 \right) \nonumber \\
    &+ d_m \left( \xi_l + \xi'_l \Delta \alpha + \frac{1}{2} \xi''_l \Delta \alpha^2 + \frac{1}{6}\xi'''_l \Delta \alpha^3 \right)\bigg], \nonumber \\
    \nonumber \\
    c = -\frac{1}{2} \mathcal{P}^{lm} &\bigg[ d_l d_m \bigg]. \label{eqn:intial_terms}
\end{align}
Importantly, $a,\, b$ and $c$ here are different from those defined in \S\ref{sec:fix_bias}.

We choose to marginalize over $\mathcal{B}$ with a uniform prior from -$\infty$ to $\infty$, as the resulting integral yields an analytic solution. Although negative values of $\mathcal{B}$ are an unphysical solution, Figure \ref{fig:L} shows that the likelihood is very sharply peaked around the true value of $\mathcal{B}$. Thus,
the unphysical region from $-\infty$ to $0$ makes a very small
contribution to the total integral.
Therefore, the result should be the same with tighter choices for the prior on $\mathcal{B}$, or if we applied a prior on $b$ instead of $\mathcal{B}$.
We tested numerically what percentage of the likelihood is in close proximity to the peak about $\mathcal{B}$. We found that 99\% of the likelihood is contained within $\pm 0.5$ of the true value of $\mathcal{B}$. 


Performing the marginalization over $\mathcal{B}$, we obtain
\begin{align}
   \mathcal{L}_{\rm marg} =  \int_{- \infty} ^{\infty} \mathcal{L}(\mathcal{B}) d\mathcal{B} = \sqrt{\frac{\pi}{a}} \exp\left(\frac{b^2}{4 a} + c\right) \label{eqn:marg_integral}
\end{align}
\noindent As in \S\ref{sec:fix_bias}, we take the log of the marginalized likelihood, differentiate with respect
to $\Delta \alpha$, and set it equal to zero.
This again yields a quadratic in terms of $\Delta \alpha$, which is then easily solved. We have
\begin{align}
    \ln \mathcal{L}_{\rm marg} = &\ln{\left(\sqrt{\frac{\pi}{a}} \exp\left(\frac{b^2}{4 a} + c\right)\right)} \nonumber \\
    = &\ln \left( \sqrt{\frac{\pi}{a}} \right) + \frac{b^2}{4 a} + c, 
\end{align}
meaning that
\begin{align}
        \frac{\partial}{\partial \Delta \alpha}[ \ln{\mathcal{L}_{\rm marg}}] = &\frac{\partial}{\partial \Delta \alpha} \left[ \ln \left( \sqrt{\frac{\pi}{a}} \right) + \frac{b^2}{4 a} + c \right] \nonumber \\
    = &- \frac{1}{2 a} \frac{\partial a}{\partial \Delta \alpha} + \frac{b}{2 a} \frac{\partial b}{\partial \Delta \alpha} - \frac{b^2}{4 a^2} \frac{\partial a}{\partial \Delta \alpha}. \label{eqn:chain_rule}
\end{align}
We want to recover a polynomial from equation (\ref{eqn:chain_rule}); to do so, we need to expand the terms in $a$, and its derivative to recover a second-degree polynomial in $\alpha$.
We Taylor expand the $a^{-1}$ and $a^{-2}$ terms in equation (\ref{eqn:chain_rule}) about $\Delta \alpha = 0$, and note that $a$ and $a^2$ are always large, so the expansion is convergent. We find these expansions about $\Delta \alpha$ = 0 are:
\begin{align}
    \frac{1}{a} \bigg|_{\Delta \alpha = 0} &= \frac{1}{c_a} - \frac{b_a}{c_a^2} \Delta \alpha + \frac{1}{2} \left( \frac{2b_a^2}{c_a^3} - \frac{2a_a}{c_a^2} \right) \Delta \alpha^2 + \cdots, \nonumber \\
     \frac{1}{a^2} \bigg|_{\Delta \alpha = 0} &= \frac{1}{c_a^2} - \frac{2b_a}{c_a^3} \Delta \alpha + \frac{1}{2} \left( \frac{6b_a^2}{c_a^4} - \frac{4a_a}{c_a^3} \right) \Delta \alpha^2 + \cdots. \label{eqn:1/a}
\end{align}

To better organize the equations and recognize what terms we are able to drop, we group the coefficients, equation (\ref{eqn:intial_terms}), with respect to $\Delta \alpha$. Since we started our process ignoring terms higher than second order in $\Delta \alpha$ in the derivative of the log likelihood, we continue to do so here in order to consistently capture all terms at second order. 

In what follows, we use a specific variable naming convention to indicate whence each term originates. A variable called $a$ will denote that it comes from a factor of $(\Delta \alpha)^2$. A variable called $b$ will denote linearity in terms of $\Delta \alpha$. Finally, a variable called $c$ arises from
the constant in the quadratic polynomial. This variable convention follows the familiar form of $a \Delta \alpha^2 + b \Delta \alpha + c$, so in an equation the dimensions in terms of $\Delta \alpha$ are easily checked. A subscript $a$ represents that that specific variable came from the $a$ coefficient of the $\mathcal{B}$ quadratic polynomial. A subscript of $b$ represents that that specific variable belongs to the linear term in the $\mathcal{B}$ quadratic polynomial, and so on. A subscript of $da$ means that the variable with that subscript came from the group of variables deriving from the partial of $a$ with respect to $\Delta \alpha$, and so on with $db$, and $dc$.

With this variable naming convention, the terms in equation (\ref{eqn:chain_rule}) may be expanded and grouped in terms of $\Delta \alpha$. This enables us to distribute the terms and easily track the order of each term in $\Delta \alpha$. We have
\begin{align}
    a = \frac{1}{2} P^{lm} &\bigg[  \xi_m \xi_l + \left( \xi_m \xi'_l + \xi'_m \xi_l\right) \Delta \alpha  \nonumber \\
    & + \left( \frac{1}{2} \xi_m \xi''_l + \xi'_m \xi'_l + \frac{1}{2} \xi''_m \xi_l\right) \Delta \alpha^2 \nonumber \\
    & + \left(\frac{1}{2} \xi'_m \xi''_l + \frac{1}{2}\xi''_m \xi'_l  + \frac{1}{6}\xi'''_m \xi_l + \frac{1}{6}\xi'''_l \xi_m \right) \Delta \alpha^3 \bigg] \nonumber \\
    &\equiv a_a \Delta \alpha^2 + b_a \Delta \alpha + c_a, \label{eqn:a}
\end{align}
where $a_a,\,b_a$ and $c_a$ are defined by the last line. 

Taking the partial of $a$ with respect to $\Delta \alpha$, we obtain
\begin{align}
     \frac{\partial a}{\partial \Delta \alpha} = \frac{1}{2} P^{lm} &\bigg[ \left( \xi_m \xi'_l + \xi'_m \xi_l \right) + \left( \xi_m \xi''_l + 2\xi'_m \xi'_l + \xi''_m \xi_l \right) \Delta \alpha \nonumber \\
     &+ \left( \frac{3}{2} \xi'_m \xi''_l + \frac{3}{2} \xi''_m \xi'_l + \frac{1}{2}\xi'''_m \xi_l + \frac{1}{2}\xi'''_l \xi_m \right) \Delta \alpha^2 \bigg]  \nonumber \\
     &\equiv a_{da} \Delta \alpha^2 + b_{da} \Delta \alpha + c_{da}. \label{eqn:da}
\end{align}
We need to both expand $b$ in order to recover a second-degree polynomial in terms of $\Delta \alpha$:
\begin{align}
     b = -\frac{1}{2} P^{lm} &\bigg[  \left( -d_l \xi_m - d_m \xi_l \right) + \left( -d_l \xi_m' - d_m \xi'_l \right) \Delta \alpha \nonumber \\
     &+ \left( -\frac{1}{2} d_l \xi''_m - \frac{1}{2} d_m \xi''_l \right) \Delta \alpha^2 \bigg], \nonumber \\
     &\equiv a_b \Delta \alpha^2 + b_b \Delta \alpha + c_b. \label{eqn:b}
\end{align}
We also need to expand the partial of $b$ with respect to $\Delta \alpha$; doing so we obtain
\begin{align}
     \frac{\partial b}{ \partial \Delta \alpha} = - \frac{1}{2} P^{lm} &\bigg[ \left( -d_l \xi'_m - d_m \xi'_l \right) + \left( -d_l \xi''_m - d_m \xi''_l \right) \Delta \alpha \nonumber \\
     &+ \frac{1}{2} \left(- d_l \xi'''_m - d_m \xi'''_l \right) \Delta \alpha^2 \bigg]  \nonumber \\
     &\equiv a_{db} \Delta \alpha^2 + b_{d b} \Delta \alpha + c_{d b}. \label{eqn:db}
\end{align}


Substituting equations (\ref{eqn:a}) through (\ref{eqn:db}) into equation (\ref{eqn:chain_rule}), we are able to distribute terms and pull out a quadratic which can then be solved to give us the optimal value of $\alpha$. We have for the derivative of the log likelihood:
\begin{align}
    \frac{\partial}{\partial \Delta \alpha} &[\ln{ \mathcal{L}_{\rm marg}}] = \nonumber \\ 
    - \frac{1}{2 a} & \frac{\partial a}{\partial \Delta \alpha} + \frac{b}{2 a} \frac{\partial b}{\partial \Delta \alpha} - \frac{b^2}{4 a^2} \frac{\partial a}{\partial \Delta \alpha} = \nonumber \\
    -\frac{1}{2} &\left[ \frac{1}{c_a} - \frac{b_a}{c_a^2} \Delta \alpha + \frac{1}{2} \left( \frac{2d_a^2}{c_a^3} - \frac{2 a_a}{c_a^2} \right) \Delta \alpha^2 \right] \nonumber \\
    &\times \left[ a_{d a} \Delta \alpha^2 + b_{d a} \Delta \alpha + c_{d a} \right] \nonumber \\ 
    + \frac{1}{2} &\bigg[ a_b \Delta \alpha^2 + b_b  \Delta \alpha + c_b \bigg] \left[ \frac{1}{c_a} - \frac{b_a}{c_a^2} \Delta \alpha + \frac{1}{2}\left( \frac{2 b_a^2}{c_a^3} - \frac{2 a_a}{c_a^2} \right) \Delta \alpha^2 \right] \nonumber \\
    &\times \left[ a_{db} \Delta \alpha^2 + b_{d b} \Delta \alpha + c_{d b} \right] \nonumber \\
    - \frac{1}{4} &\left[ -a_b^2 \Delta \alpha^4 + 2b_b a_b  \Delta \alpha^3 + \left( 2a_b c_b + b_b^2 \right) \Delta \alpha^2 + 2 b_b c_b \Delta \alpha + c_b^2 \right] \nonumber \\
    &\times \left[ \frac{1}{2} \left( \frac{6 b_a^2}{c_a^4} - \frac{4 a_a}{c_a^3} \right) \Delta \alpha^2 - \frac{2 b_a}{c_a^3} \Delta \alpha + \frac{1}{c_a^2} \right] \nonumber \\
    &\times \left[ a_{d a} \Delta \alpha^2 + b_{d a} \Delta \alpha + c_{d a} \right].
\end{align}
We can extract a quadratic equation in terms of $\Delta \alpha$:
\begin{align}
\frac{\partial}{\partial \Delta \alpha} &[\ln{ \mathcal{L}_{\rm marg}}] \nonumber \\
 = &\bigg[ - \frac{a_{d a}}{2 c_a} + \frac{b_a b_{d a}}{2 c_a^2} - \frac{c_{d a}}{4} \left( \frac{2 b_a^2}{c_a^3} - \frac{2 a_a}{ c_a^2} \right) + \frac{a_b c_{d b}}{2 c_a} + \frac{b_b b_{d b}}{2 c_a} \nonumber \\
    &- \frac{c_b b_a b_{d b}}{2 c_a^2} + \frac{c_b c_{d b}}{4} \left( \frac{2 b_a^2}{c_a^3} - \frac{2 a_a}{c_a^2} \right) - \frac{c_{d a} \left( 2 a_b c_b + b_b^2 \right)}{4 c_a^2} \nonumber \\
    &- \frac{b_b c_b b_{d a}}{2 c_a^2} + \frac{b_b c_b b_a c_{d a}}{c_a^3} - \frac{c_b^2 a_{d a}}{4 c_a^2} + \frac{c_b^2 b_a b_{d a}}{2 c_a^3}  \nonumber \\
    &- \frac{c_b^2 c_{d a}}{8} \left( \frac{6 b_a^2}{c_a^4} - \frac{4 a_a}{c_a^3} \right)
    -\frac{b_a b_b c_{d b}}{2 c_a^2} + \frac{c_b a_{db}}{2 c_a} \bigg] \left( \Delta \alpha \right)^2 \nonumber \\
    + &\bigg[ \frac{b_b c_{d b}}{2 c_a} + \frac{c_b b_{d b}}{2 c_a} - \frac{c_b b_a c_{d b}}{2 c_a^2} - \frac{b_b c_b c_{d a}}{2 c_a^2} - \frac{c_b^2 b_{d a}}{4 c_a^2} \nonumber \\
    &+ \frac{c_b^2 b_a c_{d a}}{2 c_a^3} - \frac{b_{d a}}{2 c_a} + \frac{b_a c_{d a}}{2 c_a^2} \bigg] \Delta \alpha \nonumber \\
    + &\left[ \frac{c_b c_{d b}}{2 c_a} - \frac{c_b^2 c_{d a}}{4 c_a^2} - \frac{c_{d a}}{2 c_a} \right].
    \label{eqn:marginal_quadratic}
\end{align}

\noindent With the likelihood optimization equation now written as a quadratic, it is easily solved for the optimal $\Delta \alpha$.

\begin{figure*}
    \begin{subfigure}[b]{0.49\textwidth}
        \includegraphics[width=8cm]{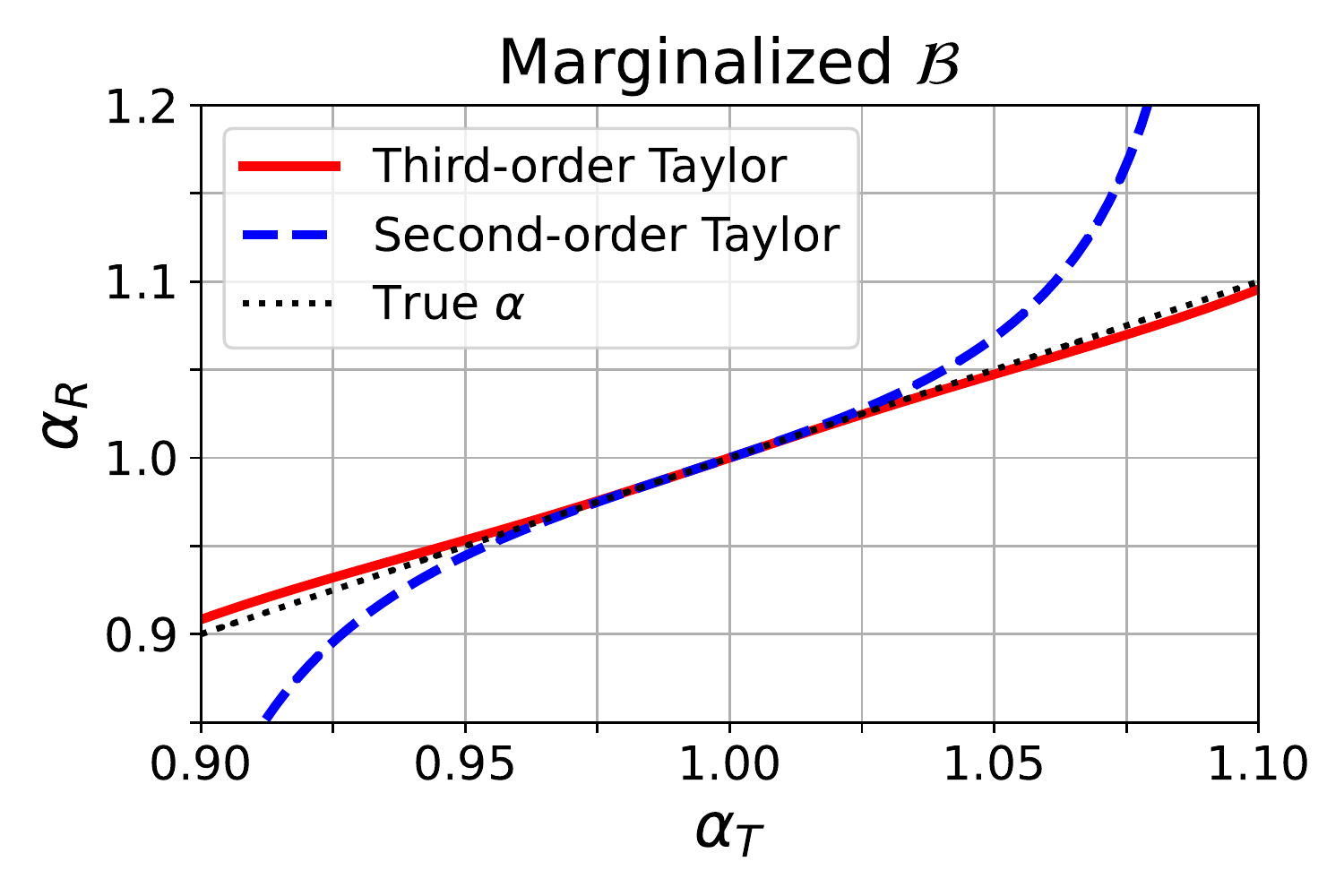}
    \end{subfigure}
    \begin{subfigure}[b]{0.49\textwidth}
        \includegraphics[width=8cm]{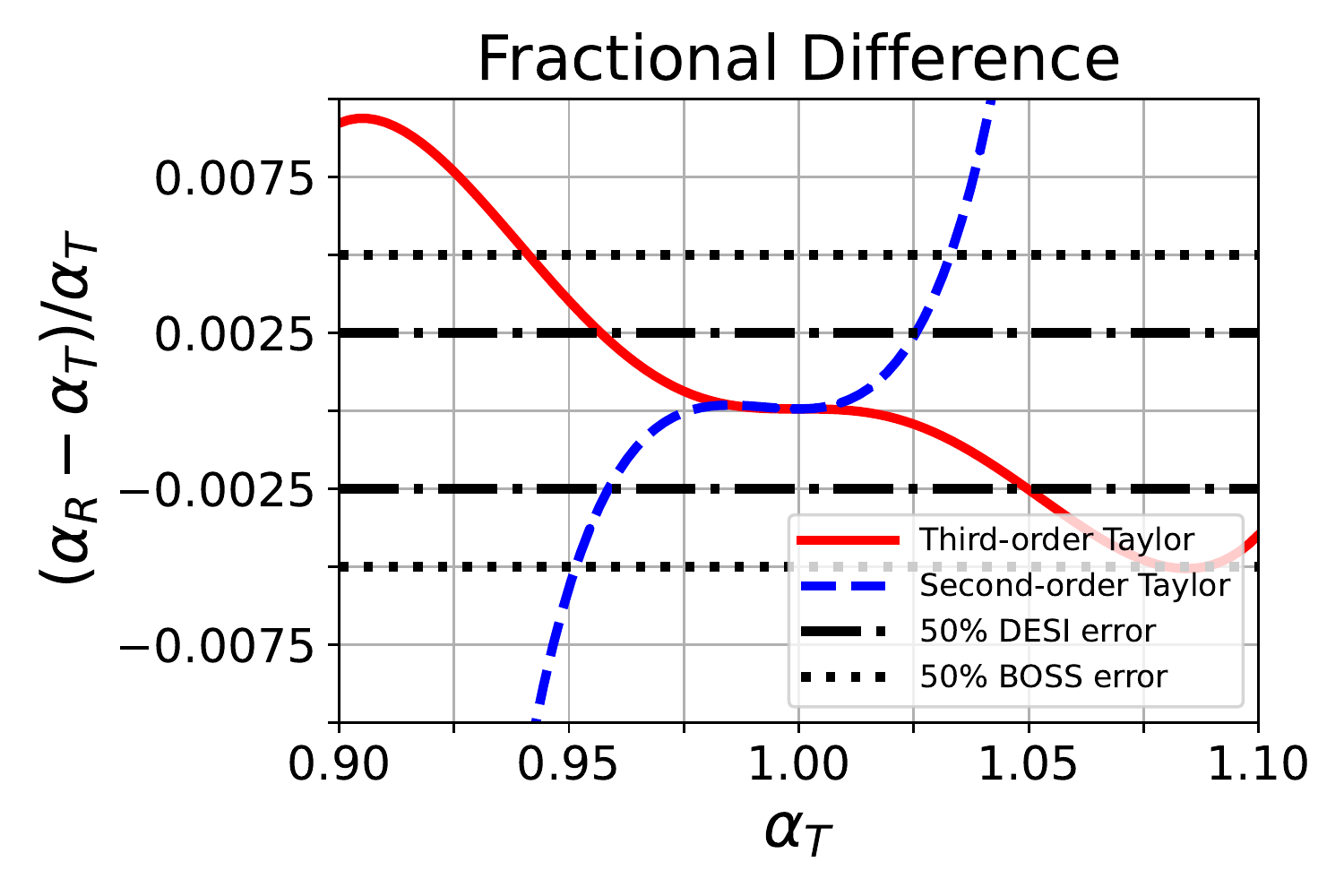}
    \end{subfigure}
    \caption{\textit{Left:} Recovered best-fit $\alpha$ of third and second-order Taylor-expanded models after marginalizing over the linear bias squared. We explored $\alpha$ in the range 0.9 to 1.1 with 100 linearly-spaced points. \textit{Right:} The fractional difference shows that our method for the third-order Taylor-expanded model is accurate up to 10\% in $\alpha$, and a smaller range in $\alpha$ for the second-order Taylor-expanded model. We show in dotted and dash-dotted lines where the error on $\alpha$ is 0.5\% and 0.25\%.  By comparison, BOSS achieves $\sim 1\%$ accuracy on $\alpha$, and DESI achieves $\sim 0.5\%$ accuracy.  This comparison shows that our method adds negligible systematic error relative to DESI's statistical precision over a range of almost 10\% in $\alpha$. As we remark in Figure \ref{fig:fixed_b1}, any DESI analysis would certainly be able to center the Taylor expansion much closer than this to whatever true $\alpha$ will then be found.}
    \label{fig:marginal_b1}
\end{figure*}

Figure~\ref{fig:marginal_b1} displays the recovered values for the optimal $\alpha$ over a range of different inputted $\alpha$. In Figure \ref{fig:marginal_b1}, for 0.5\% errors on the distance scale, our method with a third-order Taylor-expanded model works in the range of $0.94 < \alpha < 1.1$. If we demand at most 0.25\% errors in the recovered $\alpha$, our method may be employed in the range $0.954 < \alpha < 1.05$.
The range over which the third
order method works to 0.5\% or 0.25\% is not much larger than the range
over which the second order method works, but the third
order method is considerably more robust to large errors in $\alpha$ just outside of that range. Furthermore, the accuracy
achieved by the third order method is sufficient to
make only a small systematic error contribution to the DESI measurement of $\alpha$.

In a realistic analysis, one might wish to impose a more informative or restrictive prior on the linear bias or other nuisance parameters. Our method can accommodate a Gaussian or flat prior (over a finite range) on $\mathcal{B}$. The flat prior would yield an error function for $\mathcal{L}_{\rm marg}$ in equation (\ref{eqn:marg_integral}). We could then Taylor expand it, much as we Taylor-expanded $a^{-1}$ and $a^{-2}$ in equation (\ref{eqn:1/a}), and proceed with our method. Likewise, a Gaussian prior on $\mathcal{B}$ would simply lead to a re-definition of $a$ and $b$ in equation (\ref{eqn:marg_integral}), and the rest of the derivation of best-fit $\Delta \alpha$ can proceed unchanged.

Priors on linear bias $b_1$, rather than $\mathcal{B} \equiv b_1^2$, are more complicated. A Gaussian prior on $b_1$ would generate a quartic in the exponential of equation (\ref{eqn:likelihood_b})
\begin{equation}
    \mathcal{L}(b) = \exp{\left(-a b_1^4 + b b_1^2 + c b_1 + d\right)}
\end{equation}
The marginalization over $b_1$ can still be performed analytically, yielding an infinite series of Bessel functions \citep{OToole33b,OToole33}. Likewise, a flat (but infinite) prior on $b_1$ will also yield a quartic exponential but with $c = 0$. However, the corresponding integrals with a finite range, flat prior on $b_1$ cannot be performed analytically.

If a finite prior is imposed, one may wish to check whether the nuisance parameter $b_1$ or $\mathcal{B}$ is near the edge of its prior range. One could do this by solving for the global best fit of $\Delta \alpha$ and $b_1$, rather than the maximum of the marginalized likelihood for $\Delta \alpha$ alone.
This can be done by differentiating the natural log of equation (\ref{eqn:likelihood_b}) or its analog, yielding coupled non-linear equations in $b_1$/$\mathcal{B}$ and $\Delta \alpha$. These equations can then be solved iteratively to yield the maximum likelihood point.

\subsection{Impact of noisy data} \label{sec:noisy_comp}
Up to this point, we have purposely chosen not to include noise in the data. This was to determine how well our method recovers $\alpha$ in the ideal case, isolating the truncation error in our method from the random errors from noise. 
We now test the robustness of our method to noise in the correlation function by measuring whether $\alpha$ obtained in the marginalized linear bias case is significantly different when we add noise.

Using 5 test values of $\alpha$ (0.9, 0.95, 1.0, 1.05, 1.1), we
create 100 noisy mock data sets and estimate $\alpha$ with our method for each.
Table~\ref{tab:noise_comparison}
shows the average value of $\alpha$ for each noisy run and its standard deviation, compared to the noiseless value of $\alpha$.
We can compute the standard error and calculate the significance of the difference between the noisy and noiseless values of $\alpha$--these differences are not statistically significant. We conclude that our method is robust in the presence of noise.

\begin{table}
\centering
\centering
        \begin{tabular}{ccccc} 
        \hline
        Input & Avg. $\alpha_R$ & Std. deviation & $\alpha_R$ & Difference \\
        $\alpha_T$ & noisy & $\sigma(\alpha_R)$ & noiseless & \# of $\sigma$ \\
        \hline\hline
        0.90 & 0.9086 & 0.0055 & 0.9083 & 0.055 \\
        0.95 & 0.9537 & 0.0040 & 0.9534 & 0.087 \\
        1.00 & 0.9997 & 0.0047 & 1.0001 & -0.076 \\
        1.05 & 1.0475 & 0.0045 & 1.0473 & 0.036 \\
        1.10 & 1.0958 & 0.0058 & 1.0956 & 0.031 \\
        \hline
        \end{tabular}
        \caption{
        Our method recovers nearly the same results whether we use noiseless or noisy mocks, justifying the use of noiseless mocks through the paper.
         Average values of the recovered $\alpha$, $\alpha_R$, for a given value of true $\alpha$, $\alpha_T$, for data sets with and without noise. Third column gives the standard deviation of noisy $\alpha_R$ for the 100 trials for each $\alpha_T$ (\emph{i.e.} the one-$\sigma$ error on $\alpha$ in our method). Last column is the number of sigma by which one would have a systematic bias, in units of the standard deviation (column 3), \emph{i.e.} 0.5 means on average over the 100 trials a half-sigma shift.
        The shifts due to the noise bias are small enough that we do not significantly detect any of them (the detection significance is $\sqrt{100} \times$ the last column).
        }
    \label{tab:noise_comparison}
\end{table}

\begin{figure*}
    \begin{subfigure}[b]{0.49\textwidth}
        \includegraphics[width=8cm]{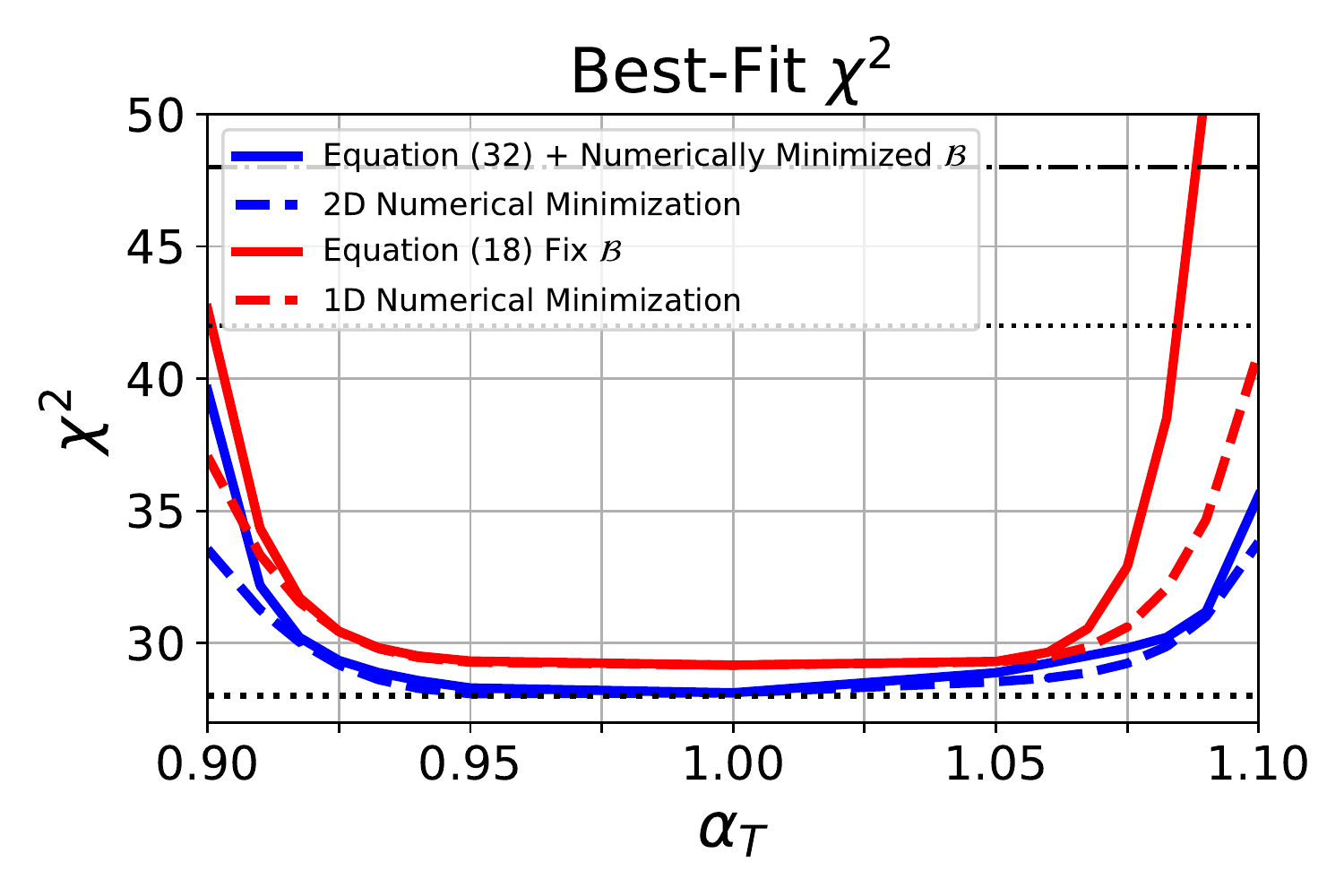}
    \end{subfigure}
    \begin{subfigure}[b]{0.49\textwidth}
        \includegraphics[width=8cm]{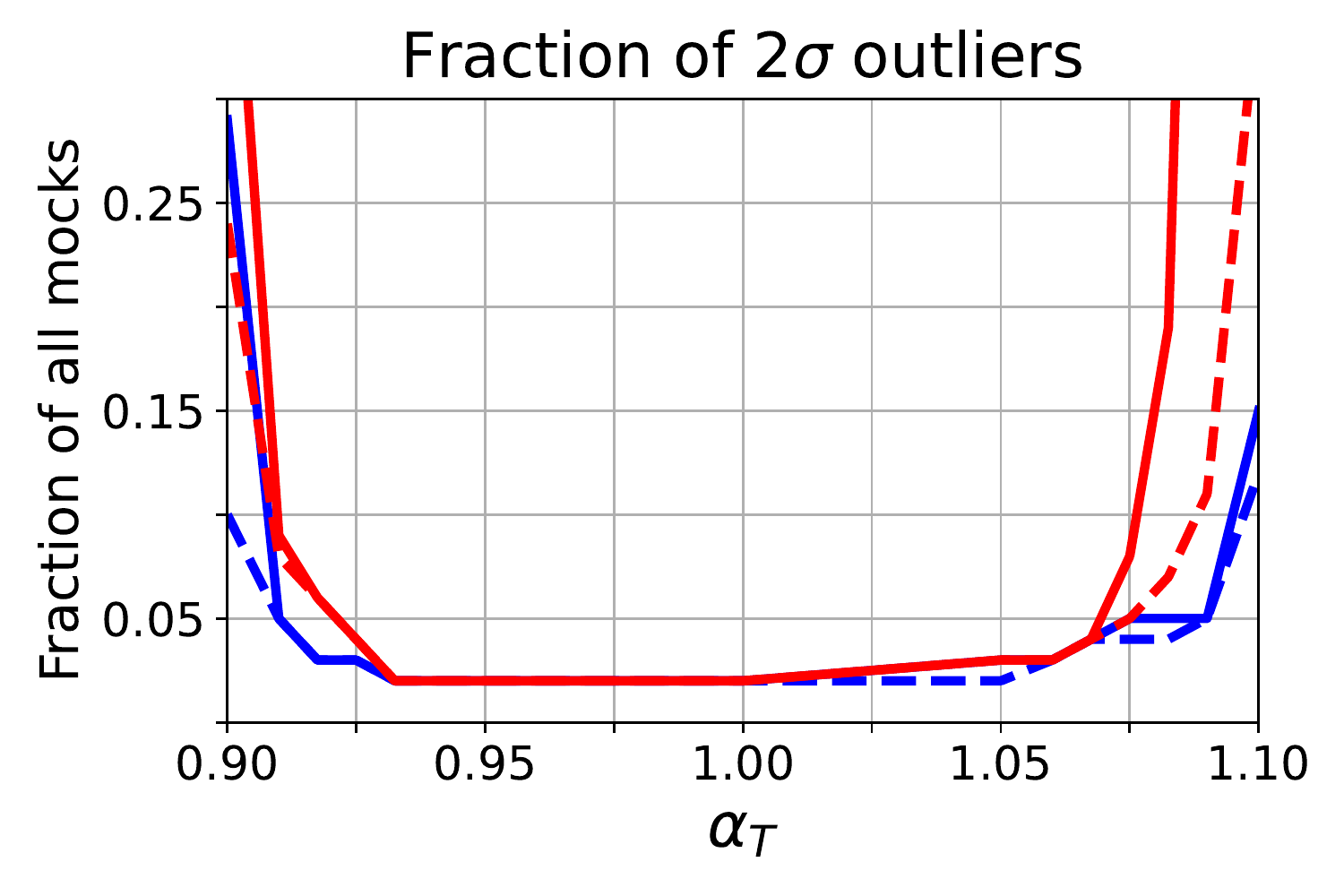}
    \end{subfigure}
    \caption{
    Goodness of fit of our third-order Taylor series model. 
    \textit{Left:} $\chi^2$ using the recovered $\alpha$ from four methods: our methods (solid) for marginalized (blue) and fixed (red) $\mathcal{B}$, and numerical minimization techniques (NM) (dashed). For the marginalized case, for the solid blue line \emph{Equation (32) + Numerically Minimized $\mathcal{B}$}, we used numerical minimization to extract the best fit $\mathcal{B}$, fixing $\alpha$ to the value recovered in equation (\ref{eqn:marginal_quadratic}). 
    This is the value of $\mathcal{B}$ that best fits the data at that fixed $\alpha$. For 2D numerical minimization, we minimized $\chi^2$ as a 2D fit over the $\alpha$ and $\mathcal{B}$ pair.  For the fixed $\mathcal{B}$ case, we compare our method from equation (\ref{eqn:18}) to a 1D numerical minimization using equation (\ref{eqn:poly_xi}) as the model.
    \textit{Right:} The fraction of mock data sets that produce more than a 2$\sigma$ deviation, as computed from the cumulative distribution function of the $\chi^2$ distribution and translated to a Gaussian-equivalent number of $\sigma$.}
    \label{fig:bestfit_chisq}
\end{figure*}

To further test the robustness of our method in the presence of noise, we measured the goodness of fit $\chi^2$ as the differences in $\Delta \alpha$ get larger. Our results are shown in Figure \ref{fig:bestfit_chisq}.
We performed and compared a series of four tests, with each test using the same methodology as the test above. 
Each test consisted of creating 100 noisy mock data sets. 
We perform two tests with fixed $\mathcal{B}$. In one case, 
we use equation (\ref{eqn:18}) to recover the best fit $\alpha$ ( labeled \emph{Equation (18) Fix $\mathcal{B}$} in Figure \ref{fig:bestfit_chisq}), and in the other, we numerically minimized the likelihood using equation (\ref{eqn:poly_xi}) as the model ( labeled \emph{1D Numerical Minimization}).
The average value of $\chi^2$ as each $\alpha$ bin is shown in Figure \ref{fig:bestfit_chisq}.

For the two cases involving a non-fixed $\mathcal{B}$, a different approach was used to create the $\chi^2$. For \emph{Equation (32) + Numerically Minimized $\mathcal{B}$}, we used equation (\ref{eqn:marginal_quadratic}) to recover the best fit $\alpha$. Then holding $\alpha$ fixed to the value we recovered, we numerically minimized the best fit value for $\mathcal{B}$. For \emph{2D Numerical Minimization}, we used numerical minimization to find the best pair of $\alpha$ and $\mathcal{B}$ for the third-order Taylor expanded model. Figure \ref{fig:bestfit_chisq} shows the results for the four tests. 

We note here the difference between using our method and the 2D numerical minimization technique is in the approximations our model makes. Equation (\ref{eqn:marginal_quadratic}) drops the higher order $\alpha^3$ and $\alpha^4$ terms as well as Taylor-expands $1/a$ and $1/a^2$. \emph{Equation(32) + Numerically Minimized $\mathcal{B}$} also carries over these approximations into retrieving the best fit value for $\mathcal{B}$ since it holds $\alpha$ fixed to the one it recovers.
As a result, the \emph{2D Numerical Minimization} obtains different values for $\alpha$ and $\mathcal{B}$, as it does not contain these approximations made in equation (\ref{eqn:likeli_compare}). Therefore, we believe that \emph{Equation(32) + Numerically Minimized $\mathcal{B}$} is the more realistic case, as it measures the goodness of fit for the best-fit $\alpha$ in our method, whereas \emph{2D Numerical Minimization} makes fewer approximations.
The difference between the two  shows that the error induced by our method is from dropping the higher order terms and approximating $1/a$ and $1/a^2$, not from the third-order model.

Figure \ref{fig:bestfit_chisq} also shows the fraction of 2$\sigma$ and 3$\sigma$ outliers in each $\alpha$ bin for each method used. If the outliers are generated by random chance, we expect 2.5 data points over a 2$\sigma$ deviation and 0.15 data points over a 3$\sigma$ deviation. We define an interval in $\alpha$ for where our model is a good fit as the interval to where there is less than double the expected random outliers. This yields an interval of $0.9325 < \alpha < 1.075$ for the fixed $\mathcal{B}$ case and $0.91 < \alpha < 1.075$ for the marginalized case. From Figure \ref{fig:fixed_b1}, the truncation errors in recovering $\alpha$ at the interval bounds are -0.002 and -0.0024 in $\Delta \alpha$, which are within half of the DESI statistical error. However, the more realistic case for the marginalized $\mathcal{B}$ correspond to $\Delta \alpha$ of 0.0085 and -0.005 in Figure \ref{fig:marginal_b1}. This is well outside our half-$\sigma$ requirement for DESI accuracy, and therefore shows that the truncation error becomes large before the model fails to provide a good fit to data.

\section{Performance Comparison}
\label{sec:performance}
\begin{table}
\centering
\centering
        \begin{tabular}{||p{3cm} p{2cm} p{2cm}||} 
        \hline
        Method & Fixed $\mathcal{B}$ ($\mu$ $\pm$ $\sigma$) & Marginalized $\mathcal{B}$ ($\mu$ $\pm$ $\sigma$) \\
        \hline\hline
        Third-order Taylor series plus analytic solution & 84 $\mu$s $\pm$ 782 ns & 170 $\mu$s $\pm$ 2.04 $\mu$s \\
        \textsc{BAOFit} & ---- & 2.04 s $\pm$ 30.6 ms \\
        Numerical minimization of splined $\xi(\alpha r)$ & 7.25 ms $\pm$ 103 $\mu$s & 8.2 ms $\pm$ 123 $\mu$s \\
        \hline
        \end{tabular}
        \caption{
        The average times to compute the optimal $\alpha$ on the three methods that are compared in the paper: our proposed method ("Third-order Taylor series plus analytic solution"), BAOFit, and a numerical minimization using \textsc{scipy}. As BAOFit does not have a comparable case to when we hold $\mathcal{B}$, there is no time for BAOFit for the "Fixed $\mathcal{B}$" case. Our method is $48\times$ faster than numerical minization over a fixed $\mathcal{B}$ and $86\times$ faster than numerical minimization using a marginalized bias. Our method is 12,000$\times$ faster than BAOFit.
        }
    \label{tab:time}
\end{table}
We compare the timing of our method to the timing of the standard method as implemented in  
the \textsc{BAOFit} package.
The method used by \textsc{BAOFit} is described in \cite{Ross15}, \cite{Anderson14} and \cite{Tojeiro14}, and has been employed in a variety of BAO studies. 
We use the \textsc{BAOFit} isotropic BAO fitting method, which also includes marginalization over polynomial broadband terms and splits the BAO feature into a wiggle piece and a no-wiggle piece.
This method analytically finds the best-fit polynomial coefficients for the broadband correlation function, and places a prior on the bias parameter, $B_{\xi}$.  The prior is a Gaussian in $\log [{B_{\xi}/B_{\rm best}}]$ with width 0.4, where $B_{\rm best}$ is the best-fit $B_{\xi}$ when $\alpha = 1$. While we discuss how to add polynomial terms to our method in Appendix \ref{sec:poly},
our method as implemented is simpler than that of \textsc{BAOFit}.
Due to this difference in methods, we also compare the timing of our method to a numerical minimization of the likelihood in equation (\ref{eqn:likeli_compare}), 
using a spline interpolation to define $\xi(\alpha r)$ rather than a Taylor series, as a more appropriate comparison.

The times for each method were measured on a 6-core Intel Core i7-8750H CPU at 2.20GHz, and reported in Table~\ref{tab:time}. The mean and standard deviation
were computed using 7 runs and 10,000 loops each for the Taylor series and the numerical minimization of splined $\xi (\alpha r)$. For \textsc{BAOFit}, the mean and standard deviation were computed using 7 runs and 10 loops each. 

With the timings shown in Table \ref{tab:time}, our method is approximately $12,000 \times$ faster than \textsc{BAOFit} (comparing the case where we marginalize over the linear bias, which is the most comparable one), $48\times$ faster than numerical minimization when marginalizing over bias, and $86\times$ faster than numerical minimization using fixed bias. Therefore our method offers significant improvements in  speed over two options one might choose as the ``standard'' methods.

\begin{table}
\centering
\centering
        \begin{tabular}{||p{3cm} p{2cm} p{2cm}||} 
        \hline
        Method & Fixed $\mathcal{B}$ & Marginalized $\mathcal{B}$ \\
        \hline\hline
        Third-order Taylor series plus analytic solution & 0.00304 & 0.00453 \\
        $\textsc{BAOFit}$ & ---- & 0.0047 \\
        Numerical minimization of splined $\xi(\alpha r)$ & 0.00321 & 0.00482 \\ \hline
        \end{tabular}
        \caption{
        The uncertainty in recovering $\alpha$ of the three methods used in this paper at an $\alpha_T$ value of 1.0. $\textsc{BAOFit}$ does not have an uncertainty for the fixed $\mathcal{B}$, since it does not have a direct comparison for that case. Our method has a smaller uncertainty than the other methods in the fixed and marginalized $\mathcal{B}$ cases.
        }
    \label{tab:uncertainty}
\end{table}

Table~\ref{tab:uncertainty} shows the uncertainty for each method with a true $\alpha$ value of 1.0. We obtained each uncertainty by finding $\alpha$ where $\chi^2$ equals its minimum plus 1, and averaging the difference between these two points and the minimum.
We chose this method since it is directly comparable to the errorbar estimation in \textsc{BAOFit}.  We used exactly the same dataset and covariance in both our method and \textsc{BAOFIT}, and made sure to match the fitting range in the correlation function (30 to 180 $h^{-1}$ Mpc).

Since our likelihood is not Gaussian, different summary statistics for the uncertainty are available: we also computed the standard deviation (Table~\ref{tab:noise_comparison}), and derived the errorbar from the 16th and 84th percentiles of the cumulative distribution function.
These different methods of calculating the uncertainty all gave similar answers.
Our method gives a comparable uncertainty to \textsc{BAOFit}.

\begin{table}
    \centering
    \begin{tabular}{l|l|c|c|c|c|c}
        \multirow{2}{*}{$z$} & \multirow{2}{*}{Sample} & Existing & DESI & Range in $\alpha$ & Lower & Upper \\
       & & $\alpha$ error & $\alpha$ error & & $\sigma$ & $\sigma$ \\
        \hline \hline
        0.15  & MGS & 3.80\% & 1.66\% & 0.92---1.13 & 2.11 & 3.42 \\ 
        0.38  & LOWZ & 1.50\% & 0.91\% & 0.94---1.07 & 4.00 & 4.67 \\ 
        0.7  & eBOSS LRG & 1.53\% & 0.48\% & 0.96---1.05 & 2.61 & 3.27 \\ 
        0.85  & eBOSS ELG & 3.25\% & 0.47\% & 0.96---1.05 & 1.23 & 1.54 \\ 
        1.48 & eBOSS QSO & 2.22\% & 0.98\% & 0.94---1.08 & 2.70 & 3.60 \\ 
        2.33 & eBOSS Ly$\alpha$ & 3.53\% & 1.09\% & 0.94---1.12 & 1.70 & 3.40 \\ 
    \end{tabular}
    \caption{
    Differences in $\alpha$ required to produce a significant systematic error for DESI.
    We find the range in $\alpha$ over which our marginalized method (Fig.~\ref{fig:marginal_b1}) produces errors <50\% the statistical error for DESI, taking the DESI forecasts at each redshift from \citet{Aghamousa16}.
    We then estimate the likelihood of such a deviation by comparing the $\alpha$ range to the existing errors on $\alpha$ from \citet{Alam21}, showing both the lower
    and upper $\sigma$ because the range in $\alpha$ is asymmetric.
    }
    \label{tab:Desi_Comp}
\end{table}

The validity of our method depends on the comparison between current and next-generation experiments, which depends on the redshift at which the measurement is taken. To test the validity of our method, we compare the accuracy of measurements at seven different redshifts from SDSS \citep{Alam21} and DESI forecasts \citep{Aghamousa16}. We show the range in $\alpha$ for which
our method with marginalized linear bias is within half of the forecasted DESI statistical error, taking this as a ceiling for allowable systematic errors.
The deviations are then converted to a number of $\sigma$ by 
dividing by the existing $\alpha$ error from SDSS; we show both the lower and upper $\sigma$ since the range in $\alpha$ can be quite asymmetric.
All errors on $\alpha$ are for the isotropic BAO scale. For DESI, they are the fractional errors on the dilation scale $R$ from Tables 2.2 and 2.3 in \citet{Aghamousa16} (rounded to the nearest redshift bin). For SDSS, we either use the isotropic error where available, or combine the errors on $D_M/r_d$ and $D_H/r_d$:
\begin{equation}
\frac{\sigma_{\alpha}}{\alpha} = \frac{\sigma_{DV}}{D_V} = \sqrt{\left(\frac{2}{3} \frac{\sigma_{DM}}{D_M}\right)^2 + \left(\frac{1}{3} \frac{\sigma_{DH}}{D_H}\right)^2}
\end{equation}
We omit the comparison at $z = 0.55$ (BOSS CMASS), because this redshift falls in between the redshift distributions for the DESI BGS and LRG samples in  \cite{Aghamousa16}, leading to a poor forecasted constraint on $\alpha$.\footnote{The DESI LRG sample \citep{Zhou22} extends to lower redshift than the LRG $N(z)$ used in \cite{Aghamousa16}, so the forecast at $z=0.55$ is unrealistically pessimistic.}

Table~\ref{tab:Desi_Comp} suggests that a 3-4$\sigma$ deviation from existing values of $\alpha$ is required to invalidate our method at $z < 0.8$, whereas a $1-3\sigma$ deviation would lead to significant systematic errors for DESI at $z > 0.8$. However, the deviations in Table~\ref{tab:Desi_Comp} assume that the distance measurements are entirely uncorrelated with each other. In any realistic model (e.g.\ flat $\Lambda$CDM or its extensions), distance is a smooth function of redshift, and hence $1\sigma$ fluctuations in $\alpha$ will be closer to the minimum error from SDSS, $\sim 1.5\%$, rather than the actual value of the $\alpha$ errorbar at that redshift.
Moreover, Type Ia Supernovae provide additional information about the distance scale, which can be quite constraining at $z < 1$ \citep{Alam21}.

\section{Discussion and Conclusions}
\label{sec:conclusions}
In this work, we have shown that we can accelerate
the time to fit the BAO dilation parameter $\alpha$ to measurements of the galaxy two-point correlation function,
thus allowing for faster measurements of the cosmic distance scale as encoded by $\alpha$.
We perform a third-order Taylor expansion about $\alpha = 1$ to allow us to analytically find the best-fit $\alpha$.
We have shown the method works both at fixed linear bias and when marginalizing over linear bias.
Our method is 12,000$\times$ faster than the standard
method, in which many values of $\alpha$ are looped over
to find the minimum $\chi^2$.
While the \textsc{BAOFit} code for the standard method is somewhat
more complicated than our fitting procedure, we note that
our method is still $48-85\times$ faster than numerically
finding the maximum likelihood using \textsc{scipy}'s numerical optimizer.

The accuracy of our method is less than the expected
statistical error on the DESI survey, ensuring that it will
contribute as a negligible systematic error. The performance
 of the third-order expansion is considerably
 better than second-order: third-order allows us to recover
 $\alpha$ to within 0.8\% across a 10\% range in $\alpha$,
 whereas second-order yields a rapidly increasing error when $|\Delta \alpha| > 0.05$.


Further work remains to be done in providing an analytic solution to the optimal scaling factor. In this paper we have simplified our model to contain only the effects of the linear bias, and have outlined the math needed to include additional polynomial terms and marginalize over them, but not implemented it. However, even with the polynomial terms, this might still be considered an incomplete model. A direction of further work might be to expand our model to contain the effects of multiple bias terms, such as the quadratic, tidal tensor \citep{McDonald09, Baldauf12}, and baryon-dark matter relative velocity biases \citep{Yoo11, Slepian18, Blazek16, SlepianF15, Beutler17, Schmidt16}. Moreover, we should modify our model to use the wiggle power spectrum rather than the infrared-resummed power spectrum in equation (\ref{eqn:power}). However, we believe there is no reason to assume the methods we present in this paper would not carry over to a more complicated model.

We note that a similar method could be developed for BAO in the 3PCF, as the relevant templates for the 3PCF already exist. \cite{Slepian17_model} presents the 3PCF model in terms of double-spherical-Bessel-function integrals from which the analytical derivative can be found. In \cite{Umeh21}, the anisotropic 3PCF is also  presented in terms of double spherical Bessel function integrals. The double-spherical-Bessel-function-integral templates could also be Taylor-expended in $\alpha$ using the same recursion relations as in our work here.

Overall, we believe the present work is already a useful step towards a complete analytic approach to finding the optimal scaling factor for BAO analyses. 

\section*{Acknowledgements}
MH would like to thank ZS and AK for allowing me to pursue this area of research and donating countless hours of their time to helping me become a more thoughtful and complete researcher. MH would also like to thank Jiamin Hou, Bob Cahn, Oliver Philcox, Kiersten Meigs, Nina Brown, Jessica Chellino and Farshad Kamalinejad for useful discussions in our weekly group meetings. ZS acknowledges useful conversations with Daniel J. Eisenstein, Lado Samushia, Hee-Jong Seo, and Stephen Portillo.

\section*{Data availability}
No new data were generated or analysed in support of this research. The code is publicly available
at \url{https://github.com/matthew3hansen/BAO_Scaling_Taylor_Series}.




\bibliographystyle{mnras}
\bibliography{main}

\appendix

\section{Incorporating Marginalization Over Polynomial Terms}
\label{sec:poly}
Often, in BAO fitting, one marginalizes over inverse power laws (or equivalently, polynomials in $1/r$) (\textit{e.g.} \citealt{Ross15}); this is intended to render the BAO constraints robust to broadband variations in the 2PCF, as might occur due to \textit{e.g.} inhomogeneities in the imaging depth of the targeting survey. Broadband effects can also arise from mis-normalization of the average number density in the survey (failure of the integral constraint). Furthermore, the radial dependence of the templates associated with both quadratic and tidal tensor biases are fairly smooth, and it is expected that marginalizing over these inverse power laws will also render the BAO fitting robust against neglecting these bias terms in the 2PCF model.

Here, we show how such marginalization can be incorporated in our framework; though the math is extensive it is conceptually simple. We leave implementation of these formulae for future work, but present them here to show that indeed in principle such an effort would be straightforward.

Similarly to our approach in \S\ref{sec:marg_bias}, we will marginalize over $\mathcal{B}$, as well as the additional polynomial terms, $a_1, a_2, a_3$. Our new model is:
\begin{align}
    \vec{m}(r) &= \mathcal{B} \left(\xi(r) + \xi'(r) \Delta \alpha + \frac{1}{2} \xi''(r) \Delta \alpha^2 + \frac{1}{6} \xi'''(r) \Delta \alpha^3 \right) \nonumber \\
    &\qquad+ \left( \frac{a_1}{r^2} + \frac{a_2}{r} + a_3 \right).
\end{align}
The next several steps follow the same approach as in \S\ref{sec:marg_bias}. We first expand out the likelihood to see the dependence of each term on $\mathcal{B}$. We have
\begin{align}
    \mathcal{L} = \exp\left[-\frac{1}{2}\chi ^2 \right] \nonumber
\end{align}
\vspace{-10pt}
\begin{align}
    = \exp \bigg[ -\frac{1}{2}\mathcal{P}^{lm} \bigg( d_l d_m - d_l &\bigg[\mathcal{B}\bigg( \xi_m + \xi'_m \Delta \alpha + \frac{1}{2} \xi''_m \Delta \alpha^2 + \frac{1}{6} \xi'''_m \Delta \alpha^3 \bigg) \nonumber \\
    &\qquad + \left( \frac{a_1}{r_m^2} + \frac{a_2}{r_m} + a_3 \right)\bigg] \nonumber
\end{align}
\vspace{-11pt}
\begin{align}
    \qquad \qquad \qquad \qquad \qquad - d_m \bigg[\mathcal{B}\bigg( & \xi_l + \xi'_l \Delta \alpha + \frac{1}{2} \xi''_l \Delta \alpha^2 + \frac{1}{6} \xi'''_l \Delta \alpha^3 \bigg)
    \nonumber \\
    &+ \left( \frac{a_1}{r_l^2} + \frac{a_2}{r_l} + a_3 \right)\bigg] \nonumber 
\end{align}
\begin{align}
    \qquad \qquad \qquad \qquad \qquad + \bigg[\mathcal{B}\bigg( &\xi_l + \xi'_l \Delta \alpha + \frac{1}{2} \xi''_l \Delta \alpha^2 + \frac{1}{6} \xi'''_l \Delta \alpha^3 \bigg) \nonumber \\
    &+ \left( \frac{a_1}{r_l^2} + \frac{a_2}{r_l} + a_3 \right)\bigg] \nonumber 
\end{align}
\begin{align}
    \qquad \qquad \qquad \qquad \qquad &\times \bigg[\mathcal{B}\left( \xi_m + \xi'_m \Delta \alpha + \frac{1}{2} \xi''_m \Delta \alpha^2 + \frac{1}{6} \xi'''_m \Delta \alpha^3 \right) \nonumber \\
    &\qquad + \left( \frac{a_1}{r_m^2} + \frac{a_2}{r_m} + a_3 \right)\bigg]\bigg)\bigg]. \label{eqn:poly_like}
\end{align}
Just as in \S\ref{sec:marg_bias}, we may factor a quadratic in $\mathcal{B}$ out of equation (\ref{eqn:poly_like}). We have
\begin{align}
    \mathcal{L}(\mathcal{B}) &= \exp\left( -a \mathcal{B}^2 + b \mathcal{B} + c\right), 
\end{align}
and analytically performing the marginalization integral with a uniform prior in $\mathcal{B}$, we find the marginalized likelihood as
\begin{align}
   \mathcal{L}_{\rm marg} = \int_{- \infty} ^{\infty} \mathcal{L}(\mathcal{B}) d\mathcal{B} &= \sqrt{\frac{\pi}{a}} \exp\left(\frac{b^2}{4 a} + c\right). \label{eqn:int_sol}
\end{align}
The coefficients are
\begin{align}
    a = \frac{1}{2} \mathcal{P}^{lm} &\bigg[ \xi_m \xi_l + \xi_m \xi'_l \Delta \alpha + \frac{1}{2} \xi_m \xi''_l \Delta \alpha^2 + \xi'_m \xi_l \Delta \alpha + \xi'_m \xi'_l \Delta \alpha^2 \nonumber \\
    &+ \frac{1}{2} \xi'_m \xi''_l \Delta \alpha^3 + \frac{1}{2}\xi''_m \xi_l \Delta \alpha^2 + \frac{1}{2} \xi''_m \xi'_l \Delta \alpha^3 + \frac{1}{6}\xi'''_m \xi_l \Delta \alpha^3 \nonumber \\
    &+ \frac{1}{6}\xi'''_l \xi_m \Delta \alpha^3 \bigg], \nonumber
\end{align}
\begin{align}
    b = -\frac{1}{2} \mathcal{P}^{lm} &\bigg[ -d_l \left( \xi_m + \xi'_m \Delta \alpha + \frac{1}{2} \xi''_m \Delta \alpha^2  + \frac{1}{6}\xi'''_m \Delta \alpha^3 \right) \nonumber \\
    &- d_m \left( \xi_l + \xi'_l \Delta \alpha + \frac{1}{2} \xi''_l \Delta \alpha^2 + \frac{1}{6}\xi'''_l \Delta \alpha^3 \right)\nonumber \\
    &+ \left( \xi_m + \xi'_m \Delta \alpha + \frac{1}{2} \xi''_m \Delta \alpha^2 + \frac{1}{6} \xi'''_m \Delta \alpha^3 \right) \times \bigg(\frac{a_1}{r_l^2} + \frac{a_2}{r_l} + a_3\bigg) \nonumber \\
    &+ \left( \xi_l + \xi'_l \Delta \alpha + \frac{1}{2} \xi''_l \Delta \alpha^2 + \frac{1}{6} \xi'''_l \Delta \alpha^3 \right) \times \bigg(\frac{a_1}{r_m^2} + \frac{a_2}{r_m} + a_3\bigg)\bigg], \nonumber
\end{align}
\begin{align}
    c = -\frac{1}{2} \mathcal{P}^{lm} & \bigg[ d_l d_m - d_m \left(\frac{a_1}{r_l^2} + \frac{a_2}{r_l} + a_3\right) \nonumber \\
    &\qquad- d_l \left(\frac{a_1}{r_m^2} + \frac{a_2}{r_m} + a_3\right) \nonumber \\
    &\qquad+ \left(\frac{a_1}{r_m^2} + \frac{a_2}{r_m} + a_3\right)\left(\frac{a_1}{r_l^2} + \frac{a_2}{r_l} + a_3\right)\bigg].
\end{align}
Moving forward, we seek now to group $a, b$, and $c$ in terms of the additional polynomial terms. With this grouping, the integrals we need to perform next become clearer. 

Since $a$ does not depend on any of the additional polynomial terms, $a$ remains unchanged throughout the grouping process. 

To deal with $b$, we define auxiliary variables as
\begin{align}
    \delta &= -\frac{1}{2} \mathcal{P}^{lm} \bigg[ -d_l \left( \xi_m + \xi'_m \Delta \alpha + \frac{1}{2} \xi''_m \Delta \alpha^2 + \frac{1}{6} \xi'''_m \Delta \alpha^3 \right) \nonumber \\
    &\qquad \qquad - d_m \left( \xi_l + \xi'_l \Delta \alpha + \frac{1}{2} \xi''_l \Delta \alpha^2 + \frac{1}{6} \xi'''_l \Delta \alpha^3 \right)\bigg], \nonumber \\
    \gamma_{a1} &= -\frac{1}{2} \mathcal{P}^{lm} \bigg[ \frac{1}{r_l^2} \left( \xi_m + \xi'_m \Delta \alpha + \frac{1}{2} \xi''_m \Delta \alpha^2 + \frac{1}{6} \xi'''_m \Delta \alpha^3 \right) \nonumber \\
    &\qquad \qquad + \frac{1}{r_m^2} \left( \xi_l + \xi'_l \Delta \alpha + \frac{1}{2} \xi''_l \Delta \alpha^2 + \frac{1}{6} \xi'''_l \Delta \alpha^3 \right)\bigg], \nonumber \\
    \gamma_{a2} &= -\frac{1}{2} \mathcal{P}^{lm} \bigg[ \frac{1}{r_l} \left( \xi_m + \xi'_m \Delta \alpha + \frac{1}{2} \xi''_m \Delta \alpha^2 + \frac{1}{6} \xi'''_m \Delta \alpha^3 \right) \nonumber \\
    &\qquad \qquad + \frac{1}{r_m} \left( \xi_l + \xi'_l \Delta \alpha + \frac{1}{2} \xi''_l \Delta \alpha^2 + \frac{1}{6} \xi'''_l \Delta \alpha^3 \right) \bigg], \nonumber \\
    \gamma_{a3} &= -\frac{1}{2} \mathcal{P}^{lm} \bigg[ \left( \xi_m + \xi'_m \Delta \alpha + \frac{1}{2} \xi''_m \Delta \alpha^2 + \frac{1}{6} \xi'''_m \Delta \alpha^3 \right) \nonumber \\
    &\qquad \qquad + \left( \xi_l + \xi'_l \Delta \alpha + \frac{1}{2} \xi''_l \Delta \alpha^2 + \frac{1}{6} \xi'''_l \Delta \alpha^3 \right) \bigg],
\end{align}
and find that
\begin{align}
    b &= \gamma_{a1} a_1 + \gamma_{a2} a_2 + \gamma_{a3} a_3 + \delta.
\end{align}
The coefficient $c$ has cross-terms with respect to the additional polynomial terms, so grouping in terms of the individual polynomial terms only is not possible. We hence require  cross-term coefficients as well; we find
\begin{align}
    c = -\frac{1}{2} \mathcal{P}^{lm} &\bigg[ \frac{a_1^2}{r_m^2 r_l^2} + \frac{a_1 a_2}{r_m^2 r_l} + \frac{a_1 a_3}{r_m^2} + \frac{a_2^2}{r_m r_l} + \frac{a_2 a_3}{r_m} + \frac{a_3 a_1}{r_l^2} \nonumber \\
    &\qquad + \frac{a_3 a_2}{r_l} + a_3^2 - \left( \frac{d_l}{r_m^2} + \frac{d_m}{r_l^2} \right) a_1 - \left( \frac{d_l}{r_m} + \frac{d_m}{r_l} \right) a_2 \nonumber \\
    &\qquad - \left( d_m + d_l \right) a_3 + d_l d_m \bigg].
\end{align}
We now rewrite equation (\ref{eqn:int_sol}) using matrices to enable us to integrate it via the solution for the $N$-dimensional Gaussian integral with a linear term:
\begin{align}
    &\int \exp \left( -\frac{1}{2} \sum_{i, j = 1}^{N} A_{ij} x_i x_j +  \sum_{i = 1}^{N} B_i x_i\right) d^N x \nonumber \\
    &\qquad\qquad = \sqrt{\frac{(2\pi)^N}{\det A}} \exp \left( \frac{1}{2} \vec{B}^T {\bf A}^{-1} \vec{B}\right).
    \label{eqn:N_gauss}
\end{align}
With this formula, equation (\ref{eqn:N_gauss}), in mind, the matrix ${\bf A}$ and the vector $\vec{B}$ are defined as:
\begin{align}
    &{\bf A} = \\
    &\begin{bmatrix}
    \frac{\gamma_{a1}^2}{4a} + \frac{1}{r_m^2 r_l^2} & \frac{2 \gamma_{a1} \gamma_{a2}}{4a} + \frac{1}{r_m^2 r_l} + \frac{1}{r_m r_l^2} & \frac{2 \gamma_{a1} \gamma_{a3}}{4a} + \frac{1}{r_m^2} + \frac{1}{r_l^2}\\
    \frac{2 \gamma_{a1} \gamma_{a2}}{4a} + \frac{1}{r_m^2 r_l} + \frac{1}{r_m r_l^2} & \frac{\gamma_{a2}^2}{4a} + \frac{1}{r_m r_l} & \frac{2 \gamma_{a2} \gamma_{a3}}{4a} + \frac{1}{r_m} + \frac{1}{r_l}\\
    \frac{2 \gamma_{a1} \gamma_{a3}}{4a} + \frac{1}{r_m^2} + \frac{1}{r_l^2} & \frac{2 \gamma_{a2} \gamma_{a3}}{4a} + \frac{1}{r_m} + \frac{1}{r_l} & \frac{\gamma_{a3}^2}{4a} + 1
    \end{bmatrix} \nonumber
\end{align}
and
\begin{align}
    &\vec{B} =  \\
    &\begin{bmatrix}
    \frac{2 \gamma_{a1} \delta}{4a} - \left( \frac{d_l}{r_m^2} + \frac{d_m}{r_l^2} \right) & \frac{2 \gamma_{a2} \delta}{4a} - \left( \frac{d_l}{r_m} + \frac{d_m}{r_l} \right) & \frac{2 \gamma_{a3} \delta}{4a} - \left( d_l + d_m \right) \nonumber
    \end{bmatrix}.
\end{align}
In our case $N = 3$ and equation (\ref{eqn:N_gauss}) gives the solution to the marginalization over each of the additional polynomial terms. We leave the final step, extremizing the resultant equation to find the optimal $\alpha$, for future work, but it is a straightforward matter of computing all of the required derivatives and solving the resulting quadratic in $\Delta \alpha$.








\bsp	
\label{lastpage}
\end{document}